\documentclass[fleqn,usenatbib]{mnras}

\usepackage{amssymb}

\usepackage{newtxtext,newtxmath}
\usepackage{bm}
\usepackage{multirow}

\usepackage[T1]{fontenc}
\usepackage[normalem]{ulem}  
\DeclareRobustCommand{\VAN}[3]{#2}
\let\VANthebibliography\thebibliography
\def\thebibliography{\DeclareRobustCommand{\VAN}[3]{##3}\VANthebibliography}

\usepackage{graphicx}	
\usepackage{amsmath}	
\usepackage{amssymb}	

\newcommand{\blue}[1]{\textcolor{blue}{#1}}

\title[Hyperon Skyrme Forces in Multi-$\Lambda$ Hypernuclei and NS Matter]{Exploring Hyperon Skyrme Forces in Multi-$\Lambda$ Hypernuclei and Neutron Star Matter}

\author[Sun et al.]{
X. D. Sun,$^{1}$\thanks{These authors contributed equally.}
S. C. Han,$^{1}$\footnotemark[1]
J. N. Hu$^{2}$
and A. Li$^{1}$\thanks{E-mail: liang@xmu.edu.cn}
\\
$^{1}$Department of Astronomy, Xiamen University, Xiamen 361005, China\\
$^{2}$School of Physics, Nankai University, Tianjin 300071, China
}

\date{Accepted XXX. Received YYY; in original form ZZZ}

\pubyear{2025}

\begin{document}
\label{firstpage}
\pagerange{\pageref{firstpage}--\pageref{lastpage}}
\maketitle

\begin{abstract}
A major source of uncertainty in modeling the strangeness-rich interiors of neutron stars arises from the poorly constrained two-body and three-body interactions among hyperons and nucleons.
We perform a comprehensive Bayesian analysis of the $\Lambda\Lambda$ and $\Lambda\Lambda N$ interaction parameters within the Skyrme Hartree-Fock framework, constrained by both hypernuclei experimental data and astrophysical observations. Our results show that the parameter space of the $\Lambda\Lambda$ interaction is tightly constrained by combining nuclear and astrophysical data, while the parameters of the $\Lambda\Lambda N$ three-body interaction remain sensitive to astrophysical inputs alone.
Specifically, the local, momentum-independent two-body interaction parameter $\lambda_0$ is tightly constrained and predominantly attractive, while the momentum-dependent parameters $\lambda_1$ and $\lambda_2$ contribute repulsive effects at high densities. 
A key role is played by the $\Lambda\Lambda$ potential depth in pure $\Lambda$ matter, which effectively constrains the two-body $\Lambda\Lambda$ interaction and governs the balance between attraction at low densities and repulsion at high densities. 
The repulsive components of $\Lambda\Lambda$ interactions then decrease hyperon fractions and 
reconcile hyperon-rich equations of state with the observed $\sim2\,M_{\odot}$ neutron stars, increasing the maximum mass by up to 22\%.
The inclusion of $\Lambda\Lambda N$ three-body forces further stiffens the EOS, raising the maximum mass by up to $\sim 0.1\,M_{\odot}$. 
Our study represents a promising step toward a complete, experimentally grounded description of dense matter across a wide range of densities and strangeness compositions.
\end{abstract}




\section{Introduction} \label{sec:intro}

Understanding the interactions between hyperons ($Y$s), as well as the possible existence and role of hyperonic three-body forces ($YNN$, $YYN$, $YYY$), is crucial for advancing our knowledge of dense matter physics, particularly in the context of neutron stars and heavy-ion collisions~\citep{2008NuPhA.804..309S,2011PhRvC..83b5804B,2021Univ....7..408L}.

While nucleon–nucleon ($NN$) and hyperon–nucleon ($YN$) interactions have been relatively well constrained by scattering experiments and hypernuclear spectroscopy, direct experimental information on hyperon–hyperon ($YY$) interactions remains extremely limited~\citep{2006PrPNP..57..564H,2007ChPhy..16.1934L,2018PhRvC..97c5206K,2020Natur.588..232A}. Current knowledge is primarily based on a handful of double-$\Lambda$ hypernuclear events and theoretical models that still suffer from considerable uncertainties~\citep{1992NuPhA.547...27D}.
Beyond two-body forces, three-body interaction play a decisive role in shaping the structure of conventional nuclei and bulk nuclear matter in \textit{ab initio} methods~\citep[e.g.,][]{2002EPJA...14..469Z,2013RvMP...85..197H,2023PhRvC.108f4316Z,2024PhRvC.109c4002W}. By analogy, hyperonic three-body forces are expected to be similarly important in hyperonic systems, particularly at high densities~\citep[e.g.,][]{1959PhRv..114..593D,2013PhRvC..87d1303L,2020EPJA...56..175G}. Their inclusion may introduce the additional repulsion required to reconcile nuclear theory with astrophysical observations—most notably, in addressing the long-standing hyperon puzzle~\citep[e.g.,][]{2016PhRvL.117r2501W,2021NPNew..31...17B}
A systematic investigation of $YY$ interactions and hyperonic three-body forces is therefore essential for constraining the equation of state (EOS) of dense matter and for guiding future experimental efforts. 
A deeper theoretical understanding of these interactions can inform the design and interpretation of upcoming hypernuclear experiments at facilities such as J-PARC, FAIR, RHIC, and Jefferson Lab, where direct constraints on hyperon forces may become accessible in future~\citep[e.g.,][]{1989PhRvC..40.2226M,1999PhRvC..59...21R}.
Insights from hypernuclear physics—particularly regarding hyperon interactions—serve as essential inputs for modeling dense matter, thereby strengthening the connection between terrestrial experiments and astrophysical phenomena. 

The present work employs Skyrme-type force models~\citep{1976AnPhy.102..226R,1981NuPhA.367..381R} to describe in-medium interactions between nucleons and hyperons~\citep[e.g.,][]{2005EPJA...24..293M,2013PhRvC..87a4333L,2015IJMPE..2450100L}, focusing specifically on the lightest hyperon, i.e., the $\Lambda$ hyperon. 
We consider three $NN$ interactions in combination with three parametrizations of the $\Lambda N$ interaction and five different choices for the $\Lambda\Lambda$ interaction.
The EOS of neutron star matter is constructed within the framework of nuclear density functional theory, enabling the calculation of key neutron star observables such as masses, radii, and tidal deformabilities. $YY$ interactions and hyperonic three-body forces have a direct impact on the stiffness of the EOS, thereby affecting these observables~\citep[e.g.,][]{2005EPJA...24..293M,2007ChPhy..16.1934L,2021Univ....7..408L,2025ApJ...982..164T,2025arXiv250912881L}, which are accessible through measurements from NICER and gravitational wave detectors.

A comprehensive analysis is carried out by combining constraints from both hypernuclear experiments and astrophysical observations, allowing us to systematically narrow down the parameter space of hyperonic interactions in dense matter.
We note that the present study aligns with our previous efforts aimed at precisely determining nuclear matter parameters near saturation density~\citep{2023ApJ...943..163Z}, quantifying the magnitude of $\Lambda N$ interaction effects~\citep{2023ApJ...942...55S}, and exploring the role of double-strangeness hypernuclei, such as those involving the $\Xi$ hyperon~\citep{2025PhRvD.112j3008D}. In these works, we have consistently pursued an integrated approach that combines nuclear experimental data and neutron star observations with modern nuclear force parametrizations to constrain the EOS of dense matter.
By linking (hyper-)nuclear measurements with neutron star observations, these studies contribute to a unified and coherent framework for understanding the physics of dense, strangeness-rich matter.

In this work, we focus on $\Lambda$ hyperons, which constitute the best-constrained strangeness sector from existing single- and double-$\Lambda$ hypernuclear data. This $\Lambda$-only treatment provides a minimal and controlled extension of nucleonic matter within the Skyrme-Hartree-Fock (SHF) framework, allowing us to isolate the impact of $\Lambda\Lambda$ and $\Lambda\Lambda N$ interactions in a Bayesian setting.
The paper is organized as follows: 
In Section 2, we outline the SHF formalism and our choices for the $NN$, $\Lambda N$, and $\Lambda\Lambda$ interactions. Section 3 details the Bayesian inference methodology. Our results are presented and discussed in Section 4, and we conclude in Section 5.

\begin{table*}
\centering
\caption{Saturation properties of symmetric nuclear matter for the Skyrme parameter sets listed in Table~\ref{tab:NN parameters}. 
$\rho_0$ is the saturation density, $E_0$ the binding energy per nucleon, $K_0$ the incompressibility, $S_0$ the symmetry energy, $L_0$ the slope of symmetry energy at saturation, and $m^*/m$ the effective mass ratio. All quantities are evaluated for symmetric nuclear matter at $\rho_0$.}
\label{tab:MR for NN}
\renewcommand\arraystretch{1.6}
\begin{tabular*}{\hsize}{@{}@{\extracolsep{\fill}}ccccccc@{}}
\hline\hline
$NN$ interaction & $\rho_0\;(\rm fm^{-3})$ & $E_0\;(\rm MeV)$ & $K_0\;(\rm MeV)$ & $S_0\;(\rm MeV)$ & $L_0\;(\rm MeV)$ & $m^*/m$ \\
\hline
SGI~\citep{1981PhLB..106..379V}   & 0.154 & -15.89 & 261.75 & 28.33 & 63.86 & 0.61 \\
SKI3~\citep{1995NuPhA.584..467R} & 0.158 & -15.98 & 258.19 & 34.82 & 100.53 & 0.58 \\
SLy4~\citep{1997NuPhA.627..710C}  & 0.160 & -15.97 & 229.91 & 32.00 & 45.94 & 0.69 \\
\hline\hline
\end{tabular*}
\end{table*}

\begin{table*}
\centering
\caption{Skyrme interaction parameter sets used in this work for $NN$, $\Lambda N$, and $\Lambda\Lambda$ forces. The parameters $\lambda_3$ and $\alpha$ characterize the density-dependent contribution of the $\Lambda\Lambda$ interaction, effectively mimicking three-body forces.
Units: $t_0$, $a_0$, $\lambda_0$ in $\rm MeV\,fm^3$; $t_1$, $a_1$, $\lambda_1$, $t_2$, $a_2$, $\lambda_2$ in $\rm MeV\,fm^5$; $t_3$, $a_3$, $\lambda_3$ in $\rm MeV\,fm^{3+3\sigma}$. 
$x_0$–$x_3$ and $\epsilon$ are dimensionless. $U_{\Lambda N}$ and $U_{\Lambda\Lambda}$ are in MeV at the specified reference density.  
}
\label{tab:NN parameters}
\renewcommand\arraystretch{1.6}
\begin{tabular*}{\hsize}{@{}@{\extracolsep{\fill}}cccccccccccccccc@{}}
\hline\hline
$NN$ interaction & $t_0$ & $t_1$ & $t_2$ & $t_3$ & $x_0$ & $x_1$ & $x_2$ & $x_3$ & $\epsilon$ \\
SGI & -1603.000 & 515.900 & 84.500 & 8000.000 & -0.020 & -0.500 & -1.731 & 0.138 & 1/3 \\
SKI3 & -1762.880 & 561.608 & -227.090 & 8106.200 & 0.308 & -1.172 & -1.091 & 1.293 & 1/4 \\
SLy4 & -2488.913 & 486.818 & -546.395 & 13777.000 & 0.834 & -0.344 & -1.000 & 1.354 & 2/3 \\
\hline
$\Lambda N$ interaction & $a_0$ & $a_1$ & $a_2$ & $a_3$ & $\gamma$ & $U_{\Lambda N}^{\rho_0}$ & - & - & - \\
YMR & -1056.2 & 26.25 & 35.00 & 1054.20 & 1/8 & -30.00 & - & - & - \\
HP$\Lambda$2 & -302.80 & 23.72 & 29.85 & 514.25 & 1 & -31.23 & - & - & - \\
SLL4 & -322.00 & 15.75 & 19.63 & 715.00 & 1 & -30.00 & - & - & - \\
\hline
$\Lambda\Lambda$ interaction & $\lambda_0$ & $\lambda_1$ & $\lambda_2$ & $\lambda_3$ & $\alpha$ & $U_{\Lambda\Lambda}^{\rho_\Lambda=0.2\rho_0}$ & - & - & - \\
SLL1 & -312.6 & 57.5 & 0.00 & 0.00 & 0.00 & -4.65 & - & - & - \\
SLL2 & -437.7 & 240.7 & 0.00 & 0.00 & 0.00 & -5.52 & - & - & - \\
SLL3 & -831.8 & 922.9 & 0.00 & 0.00 & 0.00 & -7.16 & - & - & - \\
SLL1$^\prime$ & -37.9 & 14.1 & 0.00 & 0.00 & 0.00 & -0.52 & - & - & - \\
SLL3$^\prime$ & -156.4 & 347.2 & 0.00 & 0.00 & 0.00 & -0.36 & - & - & - \\
\hline\hline
\end{tabular*}
\end{table*}

\section{Theoretical model}
\label{sec:model}

\subsection{Skyrme Hartree-Fock Formalism}
Skyrme forces~\citep{1958NucPh...9..615S} have played a central role in the development of many-body theories of nuclear structure since the early 1970s. Their parameters are typically constrained to reproduce empirical properties of finite nuclei as well as the EOS of pure neutron matter from \textit{ab initio} approaches. 
Large sets of Skyrme energy-density functionals have been systematically applied to neutron star matter, yielding neutron star mass–radius relations consistent with observational constraints~\citep{2003PhRvC..68c4324R,2012PhRvC..85c5201D}.

In the present study, we adopt standard Skyrme-type forces for $NN$ interactions, expressed in the form:
\begin{eqnarray}
V_{NN}(\bm{r}_{ij})&=&t_0(1+x_0P_{\sigma})\delta(\bm r_{ij})\nonumber\\
&\quad&+\frac{1}{2}t_1(1+x_1P_{\sigma})\left[\bm{k}_{ij}'^2\delta(\bm{r}_{ij})+\delta (\bm{r}_{ij})\bm{k}_{ij}^2\right]\nonumber\\
&\quad&+t_2(1+x_2P_{\sigma})\bm{k}_{ij}'\cdot\delta(\bm{r}_{ij})\bm{k}_{ij}\nonumber\\
&\quad&+\frac{1}{6}t_3(1+x_3P_{\sigma})\rho_N^{\epsilon}(\bm{R})\delta(\bm{r}_{ij})\nonumber\\
&\quad&+iW_0\bm{k}_{ij}'\cdot\delta(\bm{r}_{ij})(\bm{\sigma}_i+\bm{\sigma}_j)\times\bm{k}_{ij}\ , \label{eq:VNN}
\end{eqnarray}
where $\bm{r}_{ij} = \bm{r}_{i}-\bm{r}_{j}$, $\bm R =(\bm{r}_{i}+\bm{r}_{j})/2$, $\bm{k}_{ij}=-i(\overrightarrow{\nabla}_i-\overrightarrow{\nabla}j)/2$, $\bm{k}_{ij}'=i(\overleftarrow{\nabla}_i-\overleftarrow{\nabla}j)/2$. $P_{\sigma}=(1+\bm{\sigma}_i\cdot\bm{\sigma}_j)/2$ is the spin-exchange operator, and $\rho_N=\rho_n+\rho_p$ is the nucleon density.
Three widely used parameterizations: SLy4~\citep{1997NuPhA.627..710C}, SKI3~\citep{1995NuPhA.584..467R}, and SGI~\citep{1981PhLB..106..379V}, are employed here.
Their bulk properties in symmetric nuclear matter, as well as detailed parameters, are summarized in Tables~\ref{tab:MR for NN} and~\ref{tab:NN parameters}. 

The Skyrme-type interaction of a $\Lambda$ hyperon in the nuclear medium was first proposed in \citet{1981NuPhA.367..381R}, and is usually written as
\begin{eqnarray}
V_{\Lambda N}(\bm{r}_{ij})&=&u_0(1+y_0P_{\sigma})\delta(\bm r_{\Lambda N})\nonumber\\
&\quad&+\frac{1}{2}u_1\left[\bm{k}_{\Lambda N}'^2\delta(\bm{r}_{\Lambda N})+\delta (\bm{r}_{\Lambda N})\bm{k}_{\Lambda N}^2\right]\nonumber\\
&\quad&+u_2\bm{k}_{\Lambda N}'\cdot\delta(\bm{r}_{\Lambda N})\bm{k}_{\Lambda N}\nonumber\\
&\quad&+\frac{3}{8}u_3'(1+y_3P_{\sigma})\rho_N^{\gamma}\Big(\frac{\bm{r}_{N}+\bm{r}_{\Lambda}}{2} \Big)\delta(\bm{r}_{\Lambda N})\ .
\label{eq:VLambdaN}
\end{eqnarray}

For the $\Lambda\Lambda$ interaction, a Skyrme-type force in the standard form was proposed in \citet{1998PhRvC..58.3351L}, where three sets of parameters (S$\Lambda\Lambda$1, S$\Lambda\Lambda$2, and S$\Lambda\Lambda$3) were fitted to the binding energy of the double-$\Lambda$ hypernucleus $\rm ^{13}_{\Lambda\Lambda}B$. Later refinements, such as $\rm S\Lambda\Lambda1^{\prime}$ and $\rm S\Lambda\Lambda3^{\prime}$, incorporated additional constraints from actinide nuclei with two $\Lambda$ hyperons~\citep{2011NuPhA.856...55M}. The interaction reads
\begin{eqnarray}
V_{\Lambda\Lambda}(\bm{r}_{ij})&=&\lambda_0\delta(\bm r_{ij})+\frac{1}{2}\lambda_1\left[\bm{k}_{ij}'^2\delta(\bm{r}_{ij})+\delta (\bm{r}_{ij})\bm{k}_{ij}^2\right]\nonumber\\
&\quad&+\lambda_2\bm{k}_{ij}'\cdot\delta(\bm{r}_{ij})\bm{k}_{ij}+\lambda_3\rho_N^{\alpha}(\bm{R})\delta(\bm{r}_{ij})\ .
\label{eq:VLambdaLambda}
\end{eqnarray}
Here, $\lambda_0$ controls the local, momentum-independent term, $\lambda_1$ and $\lambda_2$ encode momentum dependence, while $\lambda_3$ and $\alpha$ characterize the density-dependent contribution, effectively mimicking three-body interactions~\citep{1972PhRvC...5..626V,2015IJMPE..2450100L,2005EPJA...24..293M}.

To provide a comprehensive overview of the Skyrme parameter space explored in this work, Table \ref{tab:Skyrme_parameters} summarizes the physical roles and classifications of all interaction parameters across the $NN$, $\Lambda N$, and $\Lambda\Lambda$ channels. This systematic classification highlights how each parameter contributes to the EOS through local vs. nonlocal, momentum-dependent, and density-dependent terms, offering valuable insight into the physical mechanisms governing hyperonic matter across different density regimes.

\subsection{Hamiltonian Density and EOS Construction}

The Hamiltonian density describing baryonic interactions in uniform matter can be derived from the underlying Skyrme-type potentials within the mean-field approximation (see Appendices~\ref{Appendix:A2}-\ref{Appendix:A4} for details).
For the $NN$ interaction, the Hamiltonian density is given by
\begin{eqnarray}\label{eq:Hamiltonian of NN}
    \mathcal{H}_{NN}&=&\sum_{i=n,p}\frac{\hbar^2}{2m_N}\tau_i\nonumber\\
    &\quad&+\rho_N(\tau_n+\tau_p)\Big[ \frac{t_1}{4}\Big( 1+\frac{x_1}{2} \Big)+\frac{t_2}{4}\Big( 1+\frac{x_2}{2} \Big) \Big]\nonumber\\
    &\quad&+\sum_{i=n,p}\tau_i\rho_i\Big[ -\frac{t_1}{4}\Big( \frac{1}{2}+x_1 \Big)+\frac{t_2}{4}\Big( \frac{1}{2}+x_2 \Big) \Big]\nonumber\\
    &\quad&+\frac{t_0}{2}\Big[ \Big( 1+\frac{x_0}{2} \Big)\rho_N^2-\Big( \frac{1}{2}+x_0 \Big)(\rho_n^2+\rho_p^2) \Big]\nonumber\\
    &\quad&+\frac{t_3}{12}\Big[ \Big( 1+\frac{x_3}{2} \Big)\rho_N^2-\Big( \frac{1}{2}+x_3 \Big)(\rho_n^2+\rho_p^2) \Big]\rho_N^{\epsilon} \ .
\end{eqnarray}

For the $\Lambda N$ interaction, the corresponding Hamiltonian density reads
\begin{eqnarray}
    \mathcal{H}_{\Lambda N} &=& \frac{\hbar^2}{2m_{\Lambda}}\tau_{\lambda}+u_0\Big(1+\frac{y_0}{2}\Big)\rho_N\rho_{\Lambda} \nonumber \\
    &\quad& + \frac{1}{4}(u_1+u_2)(\tau_{\Lambda}\rho_N+\tau_N\rho_{\Lambda}) \nonumber \\
    &\quad& + \frac{3}{8}u'_3\Big(1+\frac{y_3}{2}\Big)\rho^{\gamma+1}_N\rho_{\Lambda} \ .
    \label{eq:Hamiltonian of NL}
\end{eqnarray}

For the $\Lambda\Lambda$ interaction, the Hamiltonian density takes the form
\begin{eqnarray}
   \mathcal{H}_{\Lambda\Lambda}&=&\frac{\lambda_0}{4}\rho_{\Lambda}^2+\frac{1}{8}(\lambda_1+3\lambda_2)\rho_{\Lambda}\tau_{\Lambda}+\frac{\lambda_3}{4}\rho_{\Lambda}^2\rho_N^{\alpha} \ .
   \label{Eq:Hamiltonia_LL}
\end{eqnarray}

A key quantity constraining the $\Lambda \Lambda$ interaction is the potential depth in pure $\Lambda$ matter, given by
\begin{eqnarray}
    U_{\Lambda\Lambda}(\rho_{\Lambda})=\frac{\lambda_0}{2}\rho_{\Lambda}+\frac{1}{5}(3\pi^2)^{2/3}(\lambda_1+3\lambda_2)\rho_{\Lambda}^{5/3} \ . \label{Eq:well}
\end{eqnarray}

The energy density can be obtained in the Hartree-
Fock approximation with the Hamiltonian density as~\citep{1972PhRvC...5..626V,2005EPJA...24..293M}
\begin{eqnarray}
    \mathcal{E}_b =\langle \psi|\mathcal{H}|\psi \rangle  =\mathcal{E}_{NN}+\mathcal{E}_{\Lambda N}+\mathcal{E}_{\Lambda\Lambda}.
    \label{eq:energy density}
\end{eqnarray}

The effective mass of neutron $m^*_n$ is given by~\citep{1972PhRvC...5..626V,2005EPJA...24..293M}
\begin{eqnarray}
    \frac{\hbar^2}{2m^*_n}&=&\frac{\hbar^2}{2m_n}+\frac{1}{8}[t_1(2+x_1)+t_2(2+x_2)](\rho_n+\rho_p) \nonumber\\
    &\quad&-\frac{1}{8}[t_2(2x_2+1)-t_2(2x_2+1)]\rho_n \nonumber\\
    &\quad&+\frac{1}{8}[u_1(2+y_1)+u_2(2+y_2)]\rho_{\Lambda},
    \label{eq:neutron effective mass}
\end{eqnarray}
the effective mass of proton $m^*_p$ is similar to $m^*_n$ but replacing $\rho_n$ by $\rho_p$ in Eq.~(\ref{eq:neutron effective mass}), and the effective masses of $\Lambda$ hyperon can be expressed as~\citep{1981NuPhA.367..381R,1998PhRvC..58.3351L}
\begin{eqnarray}
    \frac{\hbar^2}{2m^*_\Lambda}&=&\frac{\hbar^2}{2m_\Lambda}+\frac{1}{8}[\lambda_1+3\lambda_2]\rho_{\Lambda} \nonumber\\
    &\quad&-a_1(\rho_n+\rho_p).
    \label{eq:Lambda effective mass}
\end{eqnarray}

\begin{table*}
\centering
\caption{Summary of Skyrme-type interaction parameters for $NN$, $N\Lambda$, and $\Lambda\Lambda$ channels. Each parameter is classified by its role (local vs.\ nonlocal, momentum-independent vs.\ momentum-dependent, density-dependent) and its physical contribution to the EOS.}
\label{tab:Skyrme_parameters}
\setlength{\tabcolsep}{4.8mm}
\renewcommand\arraystretch{1.2}
\begin{tabular}{lcccc}
\hline\hline
Parameter & Channel & Type & Contribution & Physical Meaning \\
\hline
$t_0, x_0$ & $NN$ & Local, momentum-independent & Central density term & Short-range attraction/repulsion \\
$t_1, x_1$ & $NN$ & Nonlocal, momentum-dependent & $\rho\tau$, $\bm{j}^2$ terms & Effective mass, kinetic correlations \\
$t_2, x_2$ & $NN$ & Nonlocal, momentum-dependent & $\rho\tau$, $\bm{j}^2$ terms & Spin-exchange, momentum asymmetry \\
$t_3, x_3, \gamma$ & $NN$ & Density-dependent & $\rho^{\gamma+2}$ term & Medium effects, saturation \\
$W_0$ & $NN$ & Spin-orbit & $\bm{J}\cdot\nabla\rho$ & Nuclear shell structure \\
\hline
$u_0, y_0$ & $N\Lambda$ & Local, momentum-independent & $\rho_N\rho_\Lambda$ & Central attraction between $N$ and $\Lambda$ \\
$u_1, u_2$ & $N\Lambda$ & Nonlocal, momentum-dependent & $\rho_N\tau_\Lambda+\rho_\Lambda\tau_N$ & Effective mass of $\Lambda$, kinetic feedback \\
$y_1, y_2$ & $N\Lambda$ & Nonlocal, momentum-dependent & Exchange terms & Spin/isospin dependence \\
$u_3, y_3, \beta$ & $N\Lambda$ & Density-dependent & $\rho_N^{\beta}\rho_\Lambda^2$ & Medium-induced repulsion \\
\hline
$\lambda_0$ & $\Lambda\Lambda$ & Local, momentum-independent & $\rho_\Lambda^2$ & Central attraction/repulsion in pure $\Lambda$ matter \\
$\lambda_1, \lambda_2$ & $\Lambda\Lambda$ & Nonlocal, momentum-dependent & $\rho_\Lambda\tau_\Lambda$, $\bm{j}_\Lambda^2$ & $\Lambda$ effective mass, dynamical corrections \\
$\lambda_3, \alpha$ & $\Lambda\Lambda$ & Density-dependent & $\rho_\Lambda^{\alpha+2}$ & Repulsive many-body effects at high density \\
\hline\hline
\end{tabular}
\end{table*}

For stellar matter, leptonic contributions from electrons and muons are included, with their energy densities given by
\begin{eqnarray}
\mathcal{E}_{l}=\frac{1}{\pi^2}\int_0^{k_l^f}k_l^2\sqrt{k_l^2+m_l^2}dk \ ,
\end{eqnarray}
where $l=e^-,\mu^-$.

The pressure follows from the thermodynamic relation,
\begin{eqnarray} \label{eq:pe}
    P(\mathcal{E})&=&\rho^2\frac{\partial}{\partial\rho}\left( \frac{\mathcal{E}_b+\mathcal{E}_l}{\rho} \right)\nonumber\\
    &=&\sum_i\rho_i\mu_i-\sum_i\mathcal{E}_i \ ,
\end{eqnarray}
with $i=n,p,\Lambda,e,\mu$, and chemical potentials defined as
\begin{eqnarray}
    \mu_i=\frac{\partial\mathcal{E}_i}{\partial\rho_i} \ .
\end{eqnarray}

The matter must satisfy baryon number conservation, charge neutrality, and $\beta$-equilibrium. The latter two conditions read
\begin{eqnarray}
\rho_p &=& \rho_e+\rho_{\mu} \ , \\
\mu_n &=& \mu_p+\mu_e \ , \nonumber\\
\mu_e &=& \mu_{\mu}\ , \nonumber\\
\mu_n+m_n &=& \mu_{\Lambda}+m_{\Lambda}\ .
\end{eqnarray}

It is noteworthy that in the SHF approach, the chemical potentials (which coincide with the single-particle energies $\epsilon_i$ at zero temperature) differ from those in RMF theory. In RMF models, the single-particle energies are defined as $E_i = \epsilon_i + m_i$, where $m_i$ is the rest mass of the particle~\citep{2006PhLB..640..150L,2016PhRvC..93a5803L}.

The uniform matter results are applicable only to the core of neutron stars. For the nonuniform crust, we construct the complete EOS by matching the pressure–energy density relation $P(\mathcal{E})$ in Eq.~(\ref{eq:pe}) to the Negele–Vautherin EOS in the medium-density regime~\citep{1973NuPhA.207..298N}, and to the Baym–Pethick–Sutherland EOS for the outer crust~\citep{1971ApJ...170..299B}.

To obtain equilibrium stellar configurations, we solve the Tolman–Oppenheimer–Volkoff equations, yielding the maximum mass $M_{\rm max}$ and the corresponding mass–radius ($M-R$) relation for direct comparison with astrophysical observations~\citep{2019ApJ...887L..21R,2021ApJ...918L..27R,2024ApJ...961...62V,2024ApJ...974..294S,2024ApJ...971L..20C}. Furthermore, to incorporate constraints from GW170817, particularly the tidal deformability measurement~\citep{2017PhRvL.119p1101A}, we compute tidal perturbations within the Schwarzschild metric framework.

The tidal deformability is characterized by the dimensionless parameter $\Lambda$, defined as 
${\Lambda}=\frac{2}{3}k_2(M/R)^{-5}$
with $k_2$ being the (quadrupolar) tidal Love number obtained from the ratio of the induced quadrupole moment $Q_{ij}$ to the external tidal field $E_{ij}$. For a binary system, the mass-weighted tidal deformability $\tilde{\Lambda}$ is given by
\begin{eqnarray}
\tilde{\Lambda} = \frac{16}{13}\frac{(m_1+12m_2)m_1^4}{(m_1+m_2)^5}\Lambda_1 + (1\leftrightarrow2) \ ,
\end{eqnarray}
which can be accurately extracted during the inspiral phase as a function of the chirp mass $\mathcal{M}=(m_1m_2)^{3/5}/(m_1+m_2)^{1/5}$.

\begin{figure}
\centering
\includegraphics[width=0.45\textwidth]{ 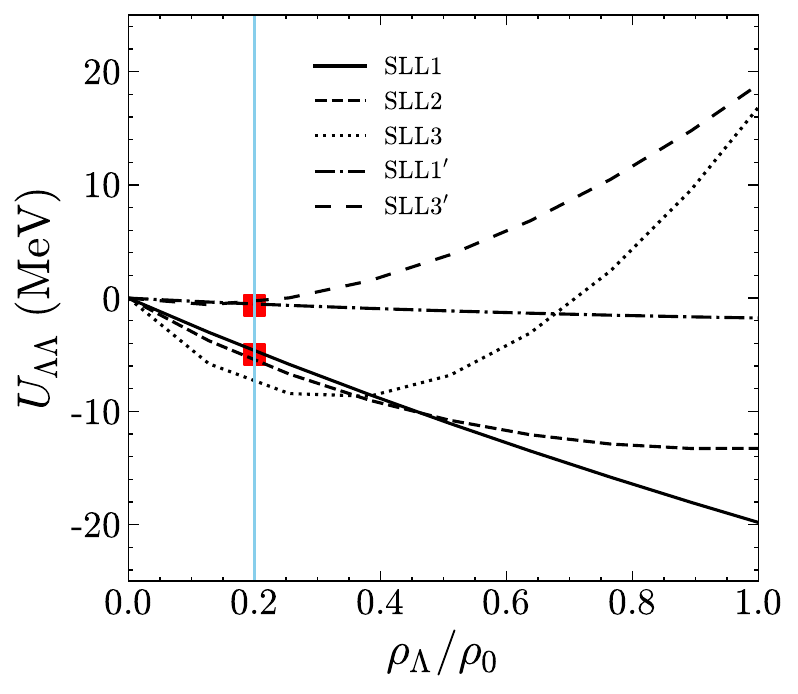}
\caption{(Colour online) The $\Lambda\Lambda$ potential depth $U_{\Lambda\Lambda}$ in pure $\Lambda$ matter as a function of the ratio of $\Lambda$ density $\rho_{\Lambda}$ to the nuclear saturation density $\rho_{0}$, calculated using the SLL1, SLL2, SLL3, SLL1$^{\prime}$, and SLL3$^{\prime}$ parameter sets of the $\Lambda\Lambda$ interaction~\citep{1998PhRvC..58.3351L,2011NuPhA.856...55M}, combined with the SLy4+SLL4 parametrization for the $\Lambda N$ sector. 
The two red squares mark the experimental values extracted from different double-$\Lambda$ hypernuclei~\citep{1995NuPhA.585...83F,2009NuPhA.828..191A,2013PhRvC..88a4003A,2015JPhG...42g5202O}, highlighting the current uncertainty in the empirical $\Lambda\Lambda$ potential depth.
}
\label{fig:U_pot}
\end{figure}
    
\begin{figure}
\centering
\includegraphics[width=0.49\textwidth]{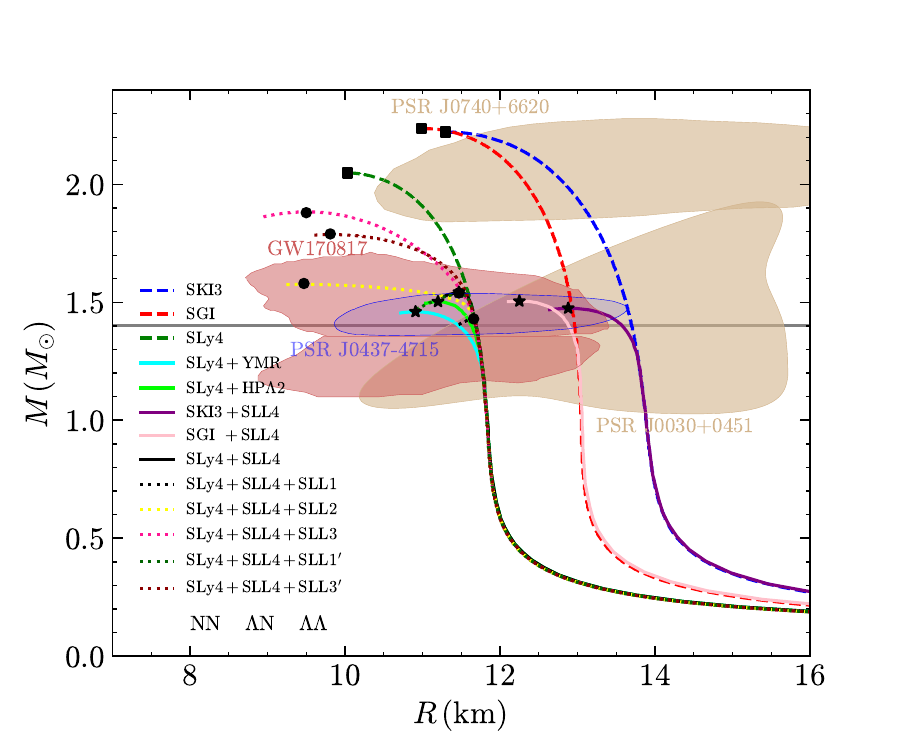}
\caption{
(Colour online) Mass–radius relations for neutron stars (dashed lines) and hyperon stars (solid/dotted lines) using different interaction combinations. Pure neutron star matter (SKI3, SGI, SLy4) is shown with black squares marking the maximum mass. Hyperon star sequences include $\Lambda N$ and $\Lambda\Lambda$ interactions, with maximum masses marked by black stars or dots—illustrating the EOS softening and maximum mass reduction that characterizes the hyperon puzzle. Observational constraints from NICER (PSR J0030+0451, PSR J0740+6620, PSR J0437–4715) and LIGO/Virgo (GW170817) are included for comparison.
}
\label{fig:TOV}
\end{figure}

\subsection{Choice of $NN$, $\Lambda N$, and $\Lambda\Lambda$ interactions}

Building upon the baryonic interactions and corresponding Hamiltonians derived in the previous section, we now specify the interaction parameter sets adopted as the foundation for the subsequent investigations.
The detailed parameters of the adopted interactions are listed in Table~\ref{tab:NN parameters}, with corresponding neutron star properties shown in Fig.~\ref{fig:TOV} (dotted lines). The full set of hyperonic star properties is presented in Table~\ref{tab:MR}.

\begin{table*}
	\centering
	\caption{Properties of neutron stars and hyperon stars for different fixed interaction combinations. The table shows maximum mass ($M_{\text{max}}$), radius around maximum mass for 2.0 $M_\odot$ stars ($R_{2.0}$), central density at maximum mass ($\rho_{\text{core}}$), critical density for $\Lambda$ hyperon appearance ($\rho_{\text{crit}}$), and properties of 1.4 $M_\odot$ stars (radius $R_{1.4}$ and tidal deformability $\Lambda_{1.4}$). These baseline results provide context for the Bayesian-constrained models presented in Section \ref{sec:result}. The table illustrates how the inclusion of hyperons softens the EOS and reduces the maximum mass, highlighting the hyperon puzzle that our Bayesian analysis aims to resolve through constrained $\Lambda \Lambda$ and $\Lambda \Lambda N$ interactions.
}
    \label{tab:MR}
\renewcommand\arraystretch{1.6}
    \begin{tabular*}{\hsize}{@{}@{\extracolsep{\fill}}ccccccccc@{}}
     \hline
      \hline
            $NN$  & $\Lambda N$ &$\Lambda\Lambda$&${M_{\rm max}/M_{\odot}}$&$R_{2.0}\;(\rm km)$ &$\rho_{\rm core}~(\rm fm^{-3})$&$\rho_{\rm crit}~(\rm fm^{-3})$&$R_{\rm 1.4}\;\rm(km)$ & $\Lambda_{1.4}$   \\
            \hline
            \multicolumn{9}{c}{\textbf{$NN$-only}} \\             
            SKI3&$-$         &$-$             &$2.22$ &$11.30$ &$0.97$ &$-$ &$13.70$&$756.64$\\
            SGI &$-$         &$-$             &$2.24$ &$10.98$ &$1.00$ &$-$ &$12.96$&$571.19$\\
            SLy4&$-$         &$-$             &$2.05$ &$10.02$ &$1.19$ &$-$ &$11.69$&$299.39$\\
            \hline
            \multicolumn{9}{c}{\textbf{$NN$+$\Lambda N$}} \\
            SLy4&YMR         &$-$             &$1.46$ &$10.90$ &$1.09$& $0.52$ &$11.47$&$263.38$\\
            SLy4&HP$\Lambda$2&$-$             &$1.54$ &$11.47$ &$0.91$ &$0.52$ &$11.69$&$285.09$\\
            SLy4&SLL4        &$-$             &$1.54$ &$11.47$ &$0.91$ &$0.52$ &$11.69$&$299.39$\\
            SKI3&SLL4        &$-$             &$1.51$ &$13.10$ &$0.71$ &$0.52$ &$13.66$&$756.64$\\
            SGI &SLL4        &$-$             &$1.54$ &$12.48$ &$0.77$ &$0.37$ &$12.48$&$571.19$\\
            \hline
            \multicolumn{9}{c}{\textbf{$NN$+$\Lambda N$+$\Lambda\Lambda$}} \\
            SLy4&SLL4        &SLL1            &$1.42$ &$11.66$ &$0.83$ &$0.52$ &$11.69$&$299.39$\\
            SLy4&SLL4        &SLL2            &$1.58$ &$9.58$  &$1.5$  &$0.52$ &$11.69$&$299.39$\\
            SLy4&SLL4        &SLL3            &$1.88$ &$9.51$  &$1.37$ &$0.52$ &$11.69$&$299.39$\\
            SLy4&SLL4        &SLL1$^{\prime}$ &$1.54$ &$11.47$  &$0.91$ &$0.52$ &$11.69$&$299.39$\\
            SLy4&SLL4        &SLL3$^{\prime}$ &$1.79$ &$9.81$  &$1.35$ &$0.52$ &$11.69$&$299.39$\\
             \hline
              \hline
	\end{tabular*}
\end{table*}

\begin{itemize}
    \item \textbf{$NN$ interaction.}
    As mentioned above, for the $NN$ sector, we employ three widely used Skyrme parameterizations: SLy4~\citep{1997NuPhA.627..710C}, SKI3~\citep{1995NuPhA.584..467R}, and SGI~\citep{1981PhLB..106..379V}. These sets are chosen primarily because (i) they reproduce reliable saturation properties of nuclear matter, and (ii) they yield neutron star properties (lines 2–4 of Table~\ref{tab:MR}) consistent with current astrophysical constraints.  

   \item \textbf{$\Lambda N$ interaction.}
    On the basis of the SLy4 $NN$ interaction, \citet{2012NuPhA.886...71G} fitted the binding energies of about 20 $\Lambda$ hypernuclei, obtaining parameter sets that reproduce the experimental data for medium and heavy hypernuclei. Among these, we adopt the best-fit parameter set HP$\Lambda$2.  
    More recently, \citet{2014PhRvC..90d7301S,2019AIPC.2130b0009S} proposed the SLL4 and SLL4$^{\prime}$ parameter sets, also based on SLy4, which reproduce the binding energies across the entire mass range of known single-$\Lambda$ hypernuclei. These were constrained by $\Lambda N$ scattering data in both light and heavy systems. For the present work, we select the SLL4 set as representative.  
    In addition, the YMR interaction, derived from the microscopic ESC08 model~\citep{2010PThPS.185...72Y}, provides a G-matrix based description of $\Lambda N$ interactions. \citet{2022EPJA...58...21C} demonstrated that both the YMR and SLL4 parametrizations yield superior agreement with experimental binding energies compared to other Skyrme-type forces. Therefore, we also employ YMR as a representative interaction.  

  For practical implementation, the parameters in Eq.~\ref{eq:VLambdaN} can be recast into effective coefficients $a_0$, $a_1$, \blue{$a_2$}, and $a_3$, defined as~\citep{2014PhRvC..90d7301S,2019AIPC.2130b0009S}:
\begin{eqnarray}
    a_0&=&u_0\Big(1+\frac{y_0}{2}\Big), \nonumber\\
    a_1&=&\frac{1}{4}(u_1+u_2), \nonumber\\
    a_2&=&\frac{1}{8}(3u_1-u_2), \nonumber\\
    a_3&=&\frac{3}{8}u'_3\Big(1+\frac{y_3}{2}\Big). 
\end{eqnarray}
  The detailed parameters of HP$\Lambda$2, SLL4, and YMR are listed in Table~\ref{tab:NN parameters}.  

   \item \textbf{$\Lambda\Lambda$ interaction and the hypernuclear constraint.}
    Our primary objective is to constrain both the two-body $\Lambda\Lambda$ and the three-body $\Lambda\Lambda N$ interactions using nuclear experimental data and astronomical observations. However, the potential depth of $\Lambda\Lambda$ interactions in pure $\Lambda$ matter at sub-saturation density remains uncertain. Specifically, at $\rho_{0}/5$ the extracted potential depth differs across experimental data: measurements of $^{10}_{\Lambda\Lambda}\rm Be$ and $^{13}_{\Lambda\Lambda}\rm B$ suggest $U^{\rm exp}_{\Lambda\Lambda}(\rho_0/5) \approx -5~\mathrm{MeV}$~\citep{1995NuPhA.585...83F}, whereas $^6_{\Lambda\Lambda}\rm He$ points to a shallower value, $U^{\rm exp}_{\Lambda\Lambda}(\rho_0/5) \approx -0.67~\mathrm{MeV}$~\citep{2009NuPhA.828..191A,2013PhRvC..88a4003A,2015JPhG...42g5202O}. These two values are compared in Fig.~\ref{fig:U_pot}.  
    In the same figure, we also include five established Skyrme-type $\Lambda\Lambda$ parameter sets (SLL1, SLL2, SLL3, SLL1$^{\prime}$, and SLL3$^{\prime}$) from earlier studies~\citep{1998PhRvC..58.3351L,2011NuPhA.856...55M}, obtained by fitting to double-$\Lambda$ hypernuclear separation energies. All sets reproduce potential depths at $\rho_0/5$ consistent with existing data.  
    We mention here that the parameters $\lambda_2$ and $\lambda_3$ in Eq.~(\ref{eq:VLambdaLambda}) were not constrained in these fits, as available experimental data involve only $s$-state double-$\Lambda$ hypernuclei. These terms correspond to $p$-wave components of the $\Lambda\Lambda$ interaction, which become relevant in multi-$\Lambda$ systems and, crucially, in neutron star matter where the $\Lambda$ population is much higher. Therefore, neutron star observables offer a unique opportunity to constrain $\lambda_2$ and $\lambda_3$.  
\end{itemize}

\begin{figure*}
\centering
\includegraphics[width=0.80\textwidth]{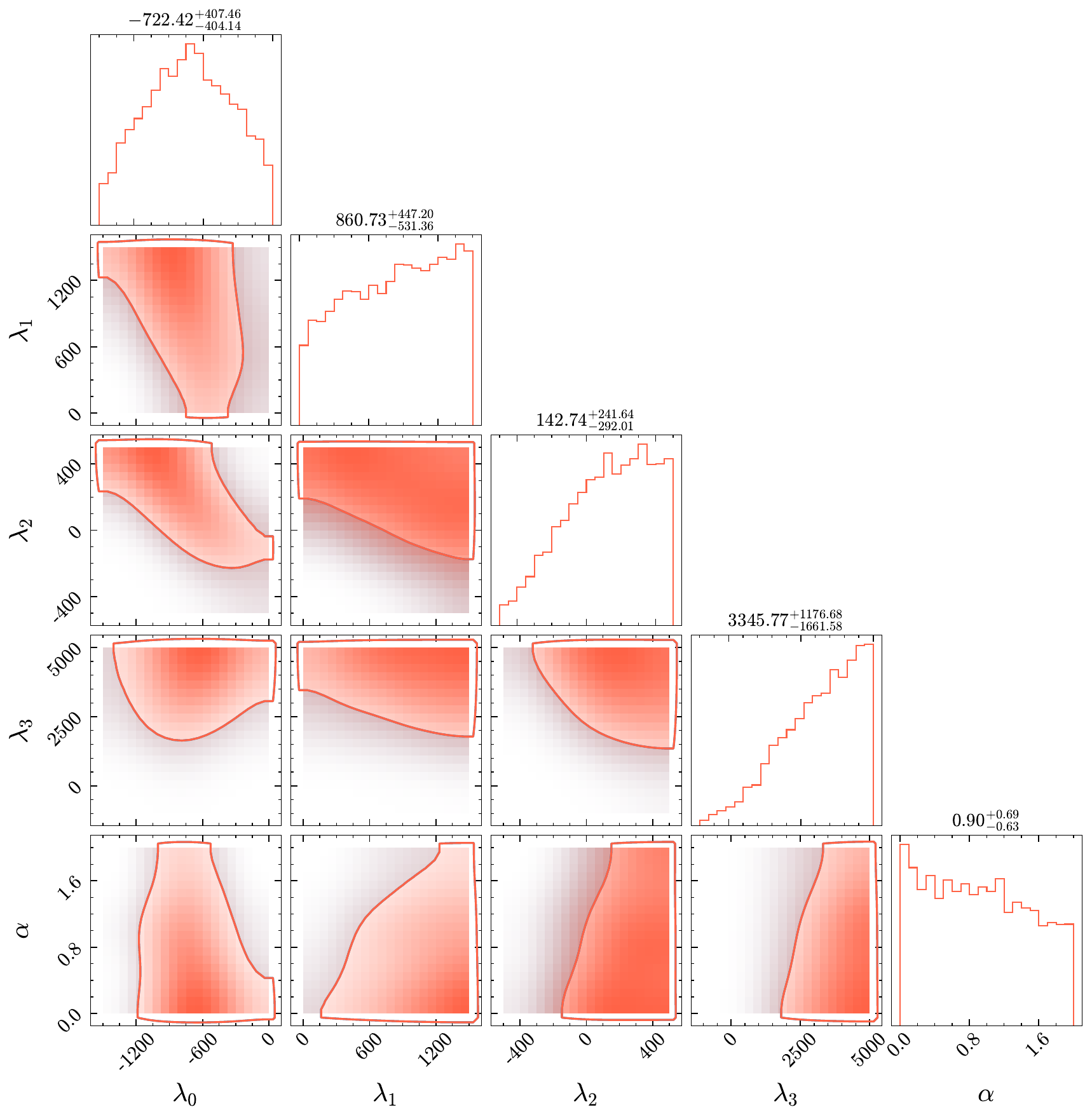}
\caption{Corner plot showing the posterior distributions and correlations for parameters$\lambda_0$, $\lambda_1$, $\lambda_2$, $\lambda_3$, $\alpha$, using the SLy4+SLL4 interaction, constrained by both astrophysical and nuclear data (+Astro+Nucl). The contours represent the 68.3\% confidence level. See Table~\ref{tab:Post_NN&NY_combined_para} for detailed numerical values.
The distributions show that $\lambda_0$ is tightly constrained and attractive, while $\lambda_0$, $\lambda_1$, and $\lambda_3$ are repulsive. The constraints exclude parameter space with low values of these parameters (lower left corners of the 2D contours).
}
\label{fig:Postpara_YY&YYN}
\end{figure*}
\begin{table*}
	\centering
	\caption{Posterior results of $\Lambda\Lambda$ and $\Lambda\Lambda$N interaction parameters based on different $NN$ and $\Lambda N$ interactions, constrained by +Astro+Nucl at 68.3\% confidence level. $\mathcal{B}$ denotes the Bayes factor relative to SLy4+YMR.
    }\label{tab:Post_NN&NY_combined_para}
\renewcommand\arraystretch{1.6}
    \begin{tabular*}{\hsize}{@{}@{\extracolsep{\fill}}cccccccc@{}}
      \hline
        \hline
			$NN$ & $\Lambda \rm N$    & $\lambda_0 ~\rm{(MeV\;fm^3)}$  &$\lambda_1 ~\rm{(MeV\;fm^5)}$ &$\lambda_2 ~\rm{(MeV\;fm^5)}$&$\lambda_3~\rm{(MeV\;fm^{3+3\alpha})}$&$\alpha$                    &$\mathcal{B}$  \\                 \hline
			SLy4 &  SLL4         & $-722.42^{+407.46}_{-404.14}$   &$860.73^{+447.20}_{-531.36}$  &$142.74^{+241.64}_{-292.01}$&$3345.77^{+1176.68}_{-1661.58}$  &$0.90^{+0.69}_{-0.63}$   &  32.23     \\
			SGI & SLL4         & $-805.90^{+394.54}_{-377.77}$  &$915.78^{+410.54}_{-541.53}$  &$197.48^{+202.89}_{-278.36}$&$2733.67^{+1536.34}_{-1790.36}$  &$0.91^{+0.73}_{-0.65}$    &  10.74      \\
			SLy4 & HP$\Lambda$2 & $-674.34^{+390.43}_{-412.72}$   &$859.62^{+438.89}_{-535.33}$  &$116.93^{+255.47}_{-305.61}$  &$3934.07^{+772.09}_{-1242.60}$  &$0.65^{+0.74}_{-0.47}$   &  8.20     \\     
			SKI3 &  SLL4         & $-819.98^{+394.23}_{-376.03}$  &$906.73^{+411.54}_{-553.15}$  &$216.70^{+192.30}_{-268.21}$&$2795.80^{+1479.59}_{-1794.91}$  &$0.94^{+0.70}_{-0.65}$    &  2.43      \\
			SLy4 &  YMR          & $-656.91^{+390.74}_{-410.91}$  &$854.82^{+448.20}_{-531.15}$  &$144.93^{+245.91}_{-329.39}$  &$4368.21^{+459.87}_{-848.45}$  &$0.45^{+0.61}_{-0.33}$  &  1.00     \\
            \hline
      Prior &  -  & U[-1500,0]  & U[0,1500]   & U[-500,500]  & U[-1000,5000]  &  U[0,2]   &   -  \\
        \hline
          \hline
	\end{tabular*}
\end{table*}

\begin{figure*}
\centering
\includegraphics[width=0.55\textwidth]{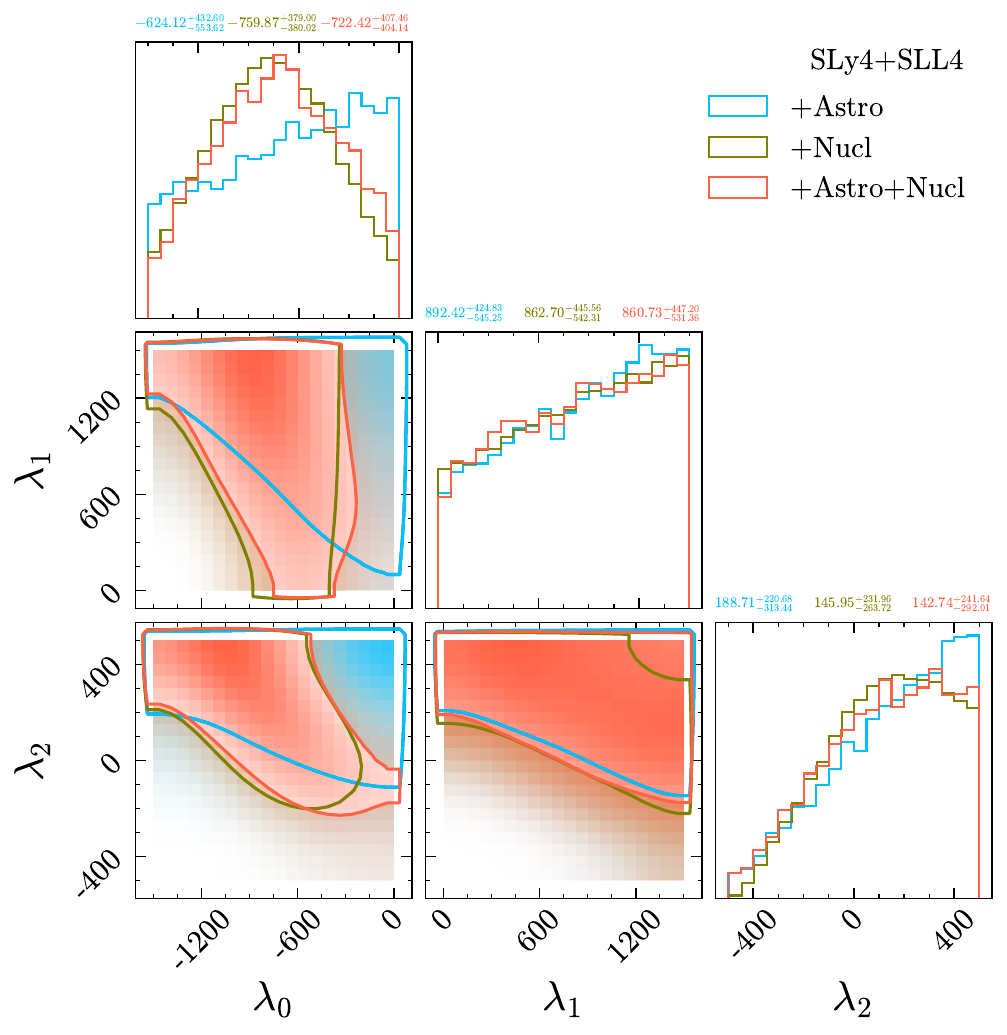}
\caption{
Posterior distributions of $\Lambda\Lambda$ parameters ($\lambda_0$, $\lambda_1$, $\lambda_2$) under different constraints: +Astro, +Nucl, and +Astro+Nucl, based on SLy4+SLL4.
}
\label{fig:Post_Astro&Nucl}
\end{figure*}
\begin{table}
	\centering
	\caption{    Posterior results of $\Lambda\Lambda$ parameters for SLy4+SLL4 under different constraints (+Astro, +Nucl, +Astro+Nucl) at 68.3\% confidence interval.
    }	\label{tab:Post_Constraints_Astro&Nucl_para}
\setlength{\tabcolsep}{0.6pt}
\renewcommand\arraystretch{1.6}
    \begin{tabular*}{\hsize}{@{}@{\extracolsep{\fill}}lcccc@{}}
    \hline\hline
			Constraint & $\lambda_0 ~\rm{(MeV \cdot fm^3)}$ &$\lambda_1 ~\rm{(MeV \cdot fm^5)}$ &$\lambda_2 ~\rm{(MeV \cdot fm^5)}$ \\
                \hline
			+Astro        & $-624.12^{+432.60}_{-553.62}$   &$892.42^{+424.83}_{-545.25}$  &$188.71^{+220.68}_{-313.44}$ \\ 
			+Nucl         & $-759.87^{+379.00}_{-380.02}$   &$862.42^{+424.83}_{-545.25}$  &$145.95^{+231.96}_{-263.72}$\\ 
			+Astro+Nucl    & $-722.42^{+407.46}_{-404.14}$   &$860.73^{+447.20}_{-531.36}$  &$142.74^{+241.64}_{-292.01}$ \\
            \hline\hline
	\end{tabular*}
\end{table}

Having constructed a complete EOS with carefully selected $NN$, $\Lambda N$, and $\Lambda\Lambda$ parameter sets that reproduce the bulk properties of nuclear matter and the empirical potential depths of $\Lambda$ hyperons, we next introduce the Bayesian analysis framework. This framework will allow us to statistically constrain hyperonic EOS models using neutron star observations.  

\section{Bayesian inference}  \label{sec:analysis} 
In Bayesian parameter estimation, the posterior distribution of a set of model parameters $\boldsymbol{\theta}$ given a dataset $\boldsymbol{D}$ is expressed as
\begin{equation}
    P(\boldsymbol\theta|\boldsymbol{D}) = \frac{P(\boldsymbol{D}|\boldsymbol\theta)P(\boldsymbol\theta)}{\int P(\boldsymbol{D}|\boldsymbol\theta)P(\boldsymbol\theta){\rm d}\boldsymbol\theta}\ ,
\end{equation}
where $P(\boldsymbol\theta)$ is the prior probability of the parameter set $\boldsymbol\theta$. The total likelihood function $P(\boldsymbol{D}|\boldsymbol\theta)$ is given by the product of the likelihoods $P_i({\boldsymbol d}_i|\boldsymbol\theta)$ associated with individual observational data ${\boldsymbol d}_i\in\boldsymbol{D}$.  
This approach has been successfully applied in several recent studies of hypernuclear matter using RMF models~\citep[e.g.,][]{2022PhRvD.106f3024M,2023ApJ...942...55S,2025PhRvD.112j3008D,2025MNRAS.536.3262H,2025PhLB..86539501L,2025arXiv251000997C}, where it has been shown to effectively integrate experimental hypernuclear data with neutron star observations.
In the following, we detail the priors and likelihoods adopted in the present analysis based on Skyrme-type energy density functionals.

\subsection{Bayesian analysis of the \texorpdfstring{$\Lambda\Lambda$}{Lambda-Lambda} interaction}

\subsubsection{Dataset and likelihood}
We consider three categories of experimental and observational data: (i) mass--radius measurements of X-ray pulsars (PSR J0030+0451, PSR J0740+6620, and PSR J0437$-$4715) from NICER, (ii) tidal deformability constraints from GW170817 observed by LIGO/Virgo, and (iii) hypernuclear data constraining the $\Lambda\Lambda$ interaction.  

\emph{Astrophysical data.}  
We include mass--radius measurements of PSR J0030+0451, PSR J0740+6620, and PSR J0437$-$4715, as well as tidal deformability from GW170817~\citep{2017PhRvL.119p1101A}. The likelihood functions $P_{\rm{NICER}}(\bm{d}_{\rm{NICER}}|\bm{\theta})$ and $P_{\rm GW}(\bm{d}_{\rm GW}|\bm{\theta})$ follow \citet{2023ApJ...942...55S}, with updated datasets: PSR J0030+0451 from \citet{2024ApJ...961...62V} replacing \citet{2019ApJ...887L..21R}, PSR J0740+6620 from \citet{2024ApJ...974..294S} replacing \citet{2021ApJ...918L..27R}, and PSR J0437$-$4715 adopting $M=1.418_{-0.037}^{+0.037}\,M_{\odot}$ and $R=11.36_{-0.95}^{+0.63}\,{\rm km}$ from \citet{2024ApJ...971L..20C}.  

\emph{Nuclear data.}  
As noted above and illustrated in Figure~\ref{fig:U_pot}, double-$\Lambda$ hypernuclear measurements from $^{10}_{\Lambda\Lambda}\rm Be$ and $^{13}_{\Lambda\Lambda}\rm Be$ suggest a potential depth of the $\Lambda\Lambda$ interaction in pure $\Lambda$ matter of $U^{\rm exp}_{\Lambda\Lambda}(\rho_0/5)\approx -5\;\rm MeV$, while $^{6}_{\Lambda\Lambda}\rm He$ points to a shallower value, $-0.67$ MeV~\citep{2015JPhG...42g5202O}. Following common practice~\citep{2015JPhG...42g5202O,2017PhRvC..95f5803F}, we adopt $U^{\rm exp}_{\Lambda\Lambda}(\rho_0/5)\approx -5\;\rm MeV$ with an experimental uncertainty $\sigma=5\;\rm MeV$, large enough to encompass the $^{6}_{\Lambda\Lambda}\rm He$ result. This ensures that the effective separation between $\Lambda$ hyperons remains consistent with empirical hypernuclear data. The likelihood function is then
\begin{equation}
    P_{\rm NUCL}({\boldsymbol d}_{\rm NUCL}|\boldsymbol\theta)=\exp \left[-\frac{\left(U^{\rm exp}_{\Lambda\Lambda}(\rho_0/5)-U^{\rm theory}_{\Lambda\Lambda}(\rho_0/5)\right)^2}{2\sigma^2}\right].
\end{equation}

\subsubsection{Model parameters and priors}\label{sec:LL_param_Prior}
The model parameters fall into three groups:

1) \emph{EOS parameters.}  
We adopt $\boldsymbol \theta_{\rm EOS} =\{\lambda_0, \lambda_1,\lambda_2, \lambda_3,\alpha\}$ for the $\Lambda\Lambda$ and $\Lambda\Lambda N$ interaction.  
Previous studies~\citep{1998PhRvC..58.3351L,2011NuPhA.856...55M} indicate $\lambda_0$ is attractive; we thus choose $\lambda_0\sim U[-1500,0]$ and $\lambda_1\sim U[0,1500]$, consistent with the parameter ranges in \citet{1998PhRvC..58.3351L,2011NuPhA.856...55M}. 
For $\lambda_2$, no direct experimental constraint exists. We treat $3\lambda_2$ in Eq.~(\ref{Eq:well}) as comparable in magnitude to $\lambda_1$, allowing both positive and negative values, with $\lambda_2\sim U[-500,500]$. The three-body $\Lambda\Lambda N$ interaction parameters $\lambda_3$ and $\alpha$ are poorly constrained; we adopt wide priors $\lambda_3\sim U[-1000,5000]$ and $\alpha\sim U[0,2]$.  

2) \emph{Central densities for NICER pulsars.}  
To evaluate the stellar mass and radius, we include the central energy density of pulsar $j$, $\varepsilon_{c,j}$, as a free parameter, yielding $M=M(\boldsymbol\theta_{\rm EOS};\varepsilon_{c,j})$ and $R=R(\boldsymbol\theta_{\rm EOS};\varepsilon_{c,j})$. We adopt $\varepsilon_{c}\sim U[0.3\times10^{15},1\times10^{15}]\,{\rm g/cm^3}$ for PSR J0030+0451, $\varepsilon_{c}\sim U[0.6\times10^{15},3\times10^{15}]\,{\rm g/cm^3}$ for PSR J0740+6620, and $\varepsilon_{c}\sim U[0.3\times10^{15},1\times10^{15}]\,{\rm g/cm^3}$ for PSR J0437$-$4715.  

3) \emph{Gravitational-wave parameters.}  
The GW data are parameterized by the chirp mass $\mathcal{M}$ and mass ratio $q$, while the component tidal deformabilities $\Lambda_{1,2}$ are determined from the EOS. We use $\mathcal{M}\sim U[1.18,1.21]\,M_\odot$ and $q\sim U[0.5,1]$.  

Sampling of the posterior distribution is performed using the python-based \textsf{bilby}~\citep{2019ascl.soft01011A} and \textsf{pymultinest}~\citep{2016ascl.soft06005B} packages.  

We conduct three main inference runs to examine the effect of various data combinations on the $\Lambda\Lambda$ interaction:  
(i) +Astro: including NICER and GW170817 constraints;  
(ii) +Nucl: including nuclear hypernuclear constraints on $U_{\Lambda\Lambda}$;  
(iii) +Astro+Nucl: combining both astrophysical and nuclear data.

\section{Results and Discussion} \label{sec:result}
We now present the posterior distributions of the hyperon interaction parameters obtained from the Bayesian analysis described in Section \ref{sec:analysis}. We first examine the constraints from the combined dataset (+Astro+Nucl) and then dissect the contributions from astrophysical and nuclear data separately.

\begin{figure*}
\centering
\includegraphics[width=0.80\textwidth]{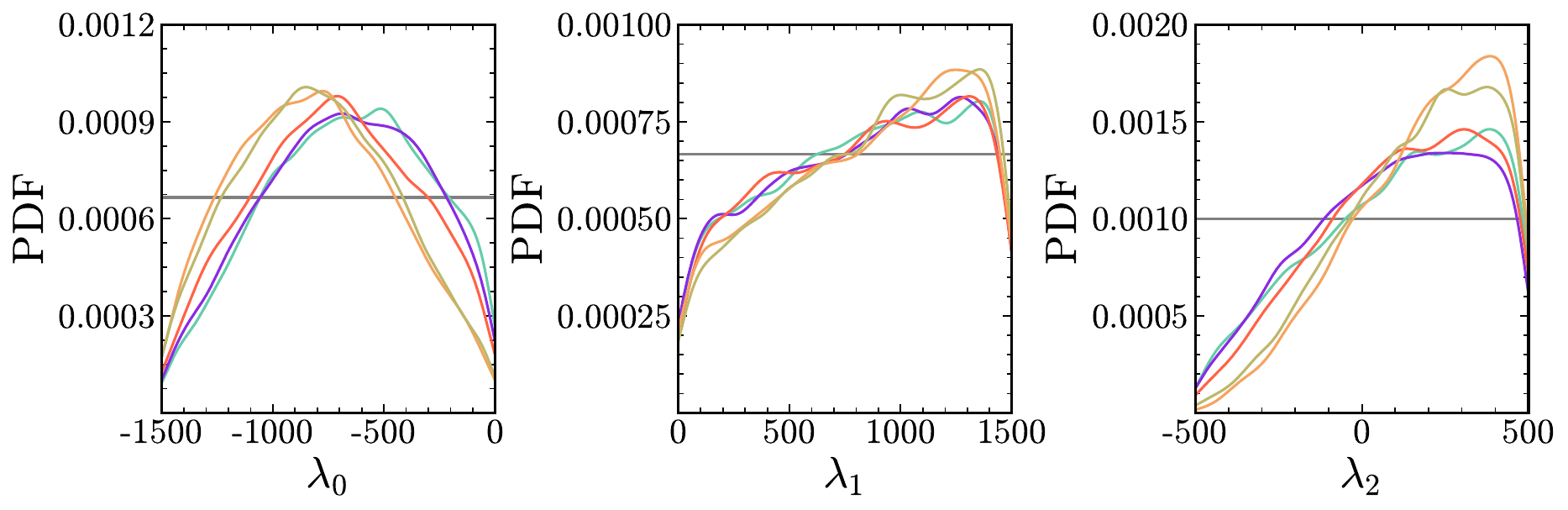}
\includegraphics[width=0.85\textwidth]{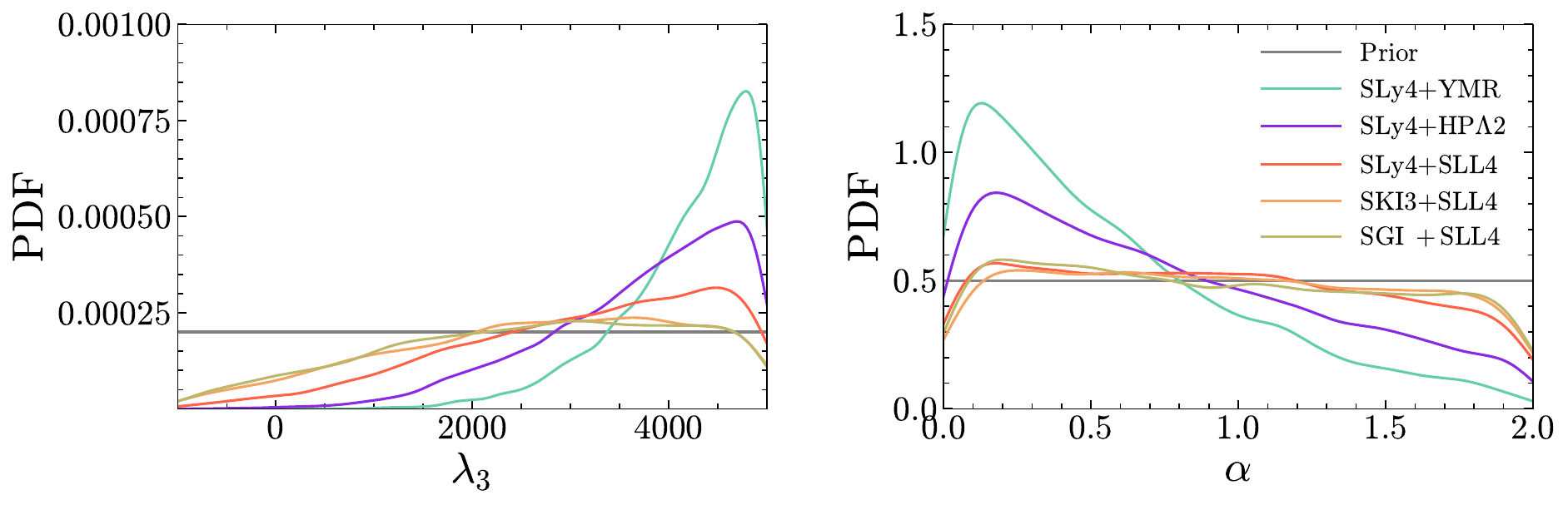}
\caption{Posterior probability distributions of $\Lambda\Lambda$ interaction parameters ($\lambda_0$, $\lambda_1$, $\lambda_2$; upper panel) and $\Lambda\Lambda$N interaction parameters ($\lambda_3$, $\alpha$; lower panel) for various $NN$+$\Lambda$ interactions, constrained by combined astrophysical and nuclear data (+Astro+Nucl).
}
\label{fig:PDF of 1D YY+YYN}
\end{figure*}

\begin{figure}
\centering
\includegraphics[width=0.45\textwidth]{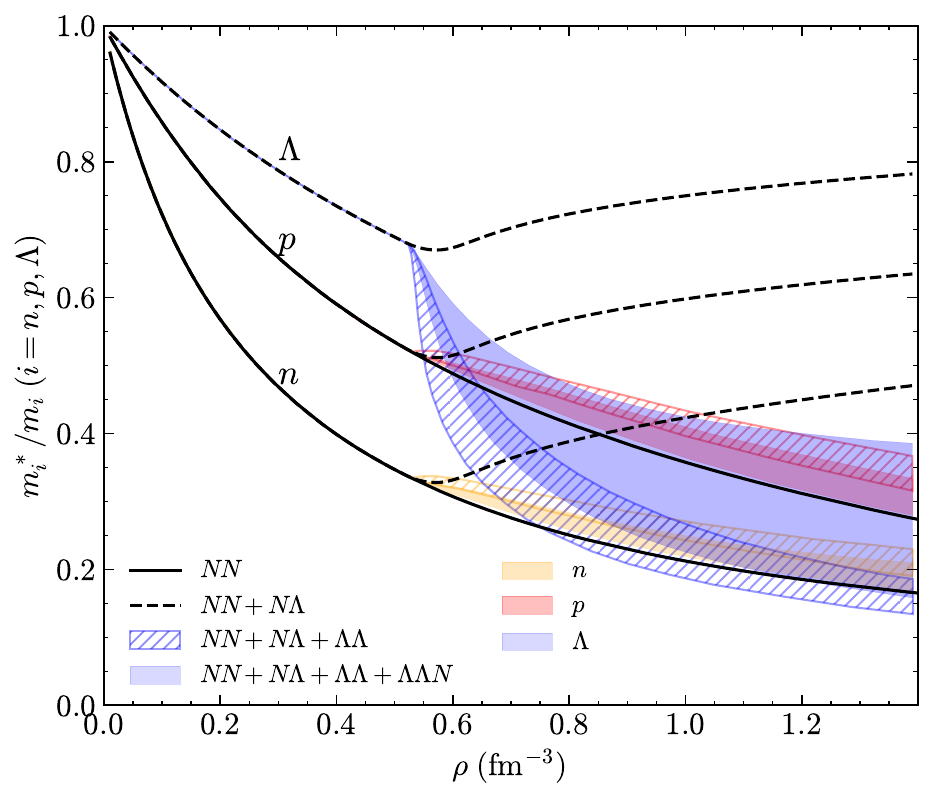}
\caption{
Posterior distributions of the effective baryon masses as functions of baryon density for the SLy4+SLL4 interaction. The black solid lines correspond to neutron star matter with $NN$ interactions only, while the black dashed lines denote hyperonic matter including $NN+\Lambda N$ interactions. The shaded regions represent the posterior distributions obtained by further including $\Lambda\Lambda$ interactions, and the shadowed regions show the results with the full $NN+\Lambda N+\Lambda\Lambda+\Lambda\Lambda N$ interaction set. The orange, red, and blue regions correspond to neutrons, protons, and $\Lambda$ hyperons, respectively.
}
\label{fig: mass of baryon}
\end{figure}

\begin{figure}
\centering
\includegraphics[width=0.49\textwidth]{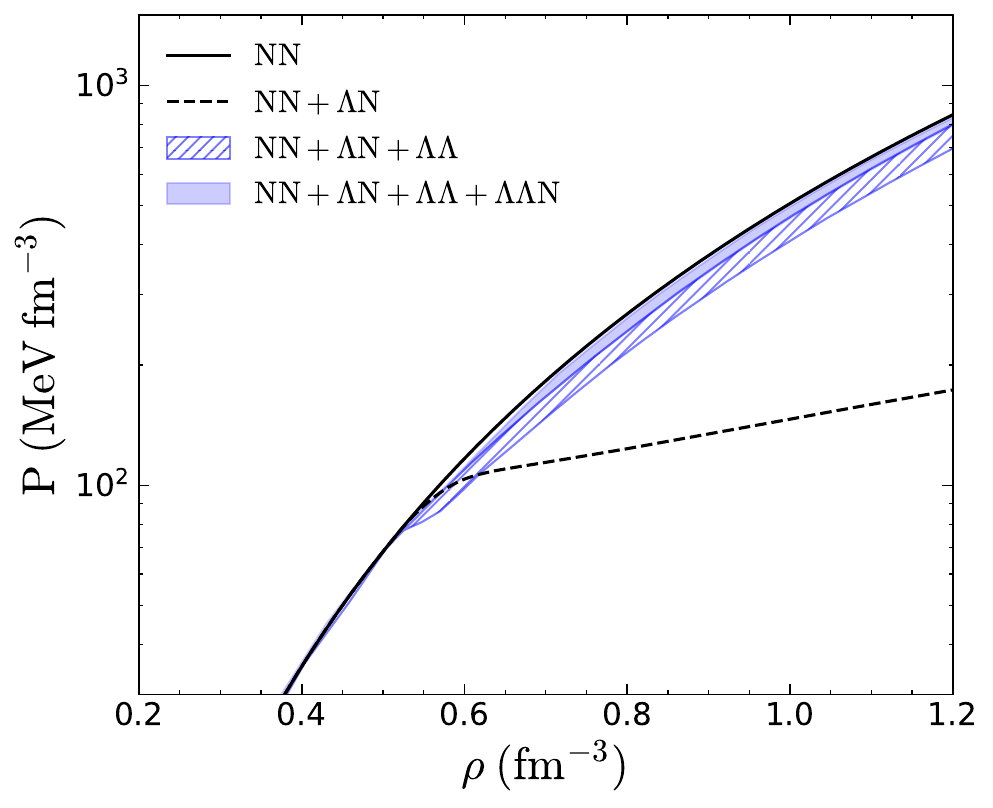}
\caption{
Posterior pressure as a function of baryon density. Black solid line: neutron star EOS ($NN$ only); black dashed: hyperon star EOS with $NN$+$\Lambda N$. Shaded regions indicate posterior distributions for $NN$+$\Lambda N$+$\Lambda\Lambda$ and $NN$+$\Lambda N$+$\Lambda\Lambda$+$\Lambda\Lambda$N, constrained by +Astro+Nucl. Representative $NN$+$\Lambda N$: SLy4+SLL4.
}
\label{fig:EOS of YY+YYN}
\end{figure}

\begin{figure*}
\centering
\includegraphics[width=0.427\textwidth]{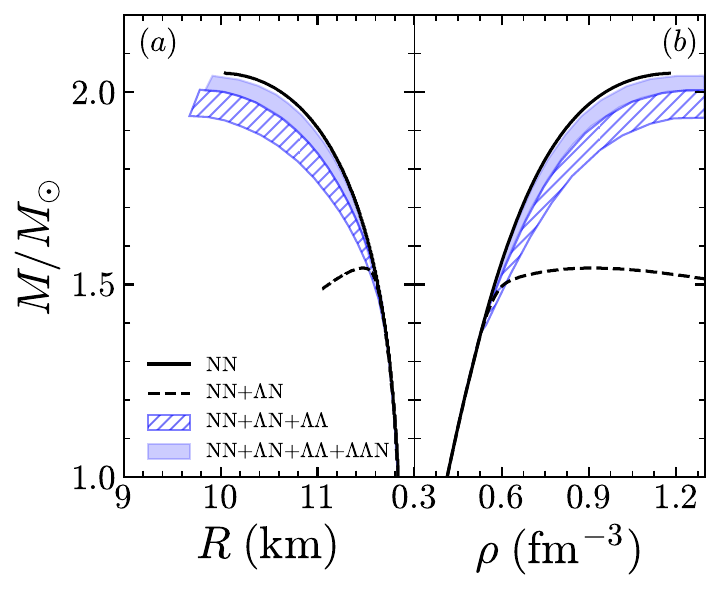}
\includegraphics[width=0.45\textwidth]{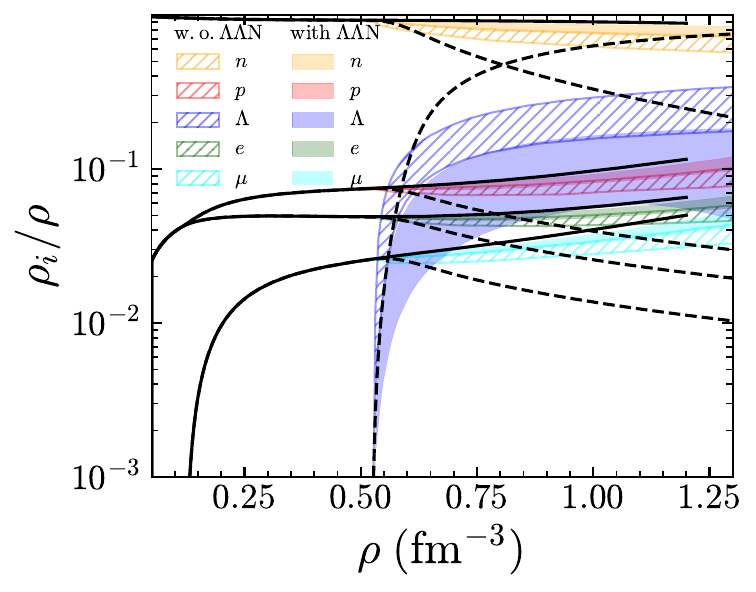}
\caption{(Left panels) Mass–radius (a) and mass–density (b) relations for different interactions, illustrating how hyperon interactions (shaded regions) stiffen the EOS relative to the softer EOS with only $\Lambda N$ forces (dashed line). (Right panel) Corresponding particle fractions as a function of baryon density, showing that repulsive hyperon interactions suppress the $\Lambda$ population at high densities. Shaded regions indicate the 68.3\% confidence intervals for hyperonic interactions, based on SLy4+SLL4 as a representative $NN$+$\Lambda N$ set. Line styles and shading follow the conventions of Figure~\ref{fig:EOS of YY+YYN}.
}
\label{fig:MR&MRrho&Density}
\end{figure*}

Figures~\ref{fig:Postpara_YY&YYN}, \ref{fig:Post_Astro&Nucl}, and \ref{fig:PDF of 1D YY+YYN} display the posterior distributions of the $\Lambda\Lambda$ and $\Lambda\Lambda N$ interaction parameters across the different physical scenarios considered. Figure~\ref{fig:Upot} illustrates the corresponding $\Lambda\Lambda$ potential depth in pure $\Lambda$ matter, which plays a central role in balancing attraction at low densities with repulsion at higher densities. 
Figure~\ref{fig: mass of baryon} shows the posterior distributions of the effective masses of neutrons, protons, and $\Lambda$ hyperons as functions of baryon density, highlighting how momentum-dependent interactions shape the kinetic contributions in dense matter.
The impact of these constraints on the dense-matter EOS and on hyperonic star structure is summarized in Figures~\ref{fig:EOS of YY+YYN} and \ref{fig:MR&MRrho&Density}, where the resulting EOS stiffness, mass–radius relations, and central densities are compared for models with and without three-body forces. 

\subsection{Posterior Distributions and Parameter Constraints} 
Figure~\ref{fig:Postpara_YY&YYN} displays the posterior distributions of $\Lambda\Lambda$ and $\Lambda\Lambda$N parameters obtained from the combined constraints of nuclear experiments and astrophysical observations (+Nucl+Astro), using the SLy4+SLL4 parameter set as our representative $NN$+$\Lambda N$ interaction. Results for alternative $NN$+$\Lambda N$ combinations are qualitatively similar and are provided in Appendix~\ref{Appendix:A1}, with the corresponding data summarized in Table~\ref{tab:Post_NN&NY_combined_para}.  

We first focus on the $\Lambda\Lambda$ parameters $\lambda_0$, $\lambda_1$, and $\lambda_2$. In the SHF framework, $\lambda_0$ represents the local momentum-independent interaction, while $\lambda_2$ and $\lambda_3$ correspond to nonlocal momentum-dependent terms. The combined nuclear and astrophysical constraints impose a tight Gaussian posterior on $\lambda_0$, peaking at $-722.42\;\mathrm{MeV\,fm^3}$. Although $\lambda_2$ and $\lambda_3$ do not follow a purely Gaussian form, their posteriors favor relatively large positive values. These results indicate that the local momentum-independent channel ($\lambda_0$) is dominated by attractive contributions, whereas the nonlocal momentum-dependent channels ($\lambda_2$, $\lambda_3$) are primarily repulsive. In particular, the parameter space at low values of $\lambda_0$, $\lambda_1$, and $\lambda_2$ is excluded, as shown in the contour maps in Figure~\ref{fig:Postpara_YY&YYN}.  

Since the $\Lambda\Lambda$ potential depth, calculated from Eq.~(\ref{Eq:well}), is defined in pure $\Lambda$ matter, it cannot constrain the three-body $\Lambda\Lambda$N force, where nucleon density vanishes. Consequently, the $\Lambda\Lambda$N parameters $\lambda_3$ and $\alpha$ are constrained exclusively by astrophysical observations. The posterior distribution shows that $\lambda_3$ disfavors low values, while $\alpha$ tends toward smaller values. Physically, larger $\lambda_3$ enhances EOS stiffness, necessary to support $2.0\,M_\odot$ stars, whereas smaller $\alpha$ corresponds to a softer density dependence, consistent with astrophysical preferences at high densities.  

Correlations among the model parameters reveal linear trends between $\lambda_0$ and $\lambda_1$, as well as between $\lambda_0$ and $\lambda_2$. These correlations merit further investigation. 
Figure~\ref{fig:Post_Astro&Nucl} presents the posterior distributions of the $\Lambda\Lambda$ interaction parameters $\lambda_0$, $\lambda_1$, and $\lambda_2$ under the separate constraints of +Astro, +Nucl, and +Astro+Nucl, based on the SLy4+SLL4 parameter set at the 68.3\% confidence level. The corresponding quantitative results are listed in Table~\ref{tab:Post_Constraints_Astro&Nucl_para}. 
It is clear that the correlations among $\lambda_0$, $\lambda_1$, and $\lambda_2$ originate primarily from nuclear experimental constraints.
Astrophysical observations favor less negative values of $\lambda_0$, implying a weaker $\Lambda\Lambda$ attraction and thus a stiffer EOS for hyperonic matter. In contrast, constraints from the $\Lambda\Lambda$ potential depth shift the posterior distribution of $\lambda_0$ toward lower values, underscoring the dominant role of nuclear experimental inputs. 
The negative correlation between $\lambda_0$ and $\lambda_1$ arises because both parameters influence the overall potential depth $U_{\Lambda\Lambda}(\rho_{\Lambda})$; a more attractive $\lambda_0$ can be compensated by a more repulsive $\lambda_1$ to maintain consistency with the hypernuclear constraint on $U_{\Lambda\Lambda}$ at $\rho_0/5$.

Figures~\ref{fig:Postpara_YY&YYN} and~\ref{fig:Post_Astro&Nucl} summarize the posterior results using SLy4+SLL4 as a representative interaction set.
To systematically examine the impact of different $NN$ and $\Lambda N$ interactions, Figure~\ref{fig:PDF of 1D YY+YYN} shows the one-dimensional posterior distribution functions (PDFs). Gray lines represent uniform priors, while colored lines represent posteriors for $\lambda_0$--$\lambda_3$ and $\alpha$ under +Astro+Nucl constraints at the 68.3\% level. For comparison, SLy4+SLL4 is taken as the baseline, while variations are introduced by replacing the $NN$ set SLy4 with SKI3/SGI or substituting the $\Lambda N$ set SLL4 with HP$\Lambda$2/YMR.
The posterior results from Figure~\ref{fig:PDF of 1D YY+YYN} are listed in Table~\ref{tab:Post_NN&NY_combined_para}, while the corresponding bulk properties of hyperon stars appear in Table~\ref{tab:Post_NN&NY_combined_MR} of Appendix \ref{Appendix:A1}.

\begin{figure}
\centering
\includegraphics[width=0.45\textwidth]{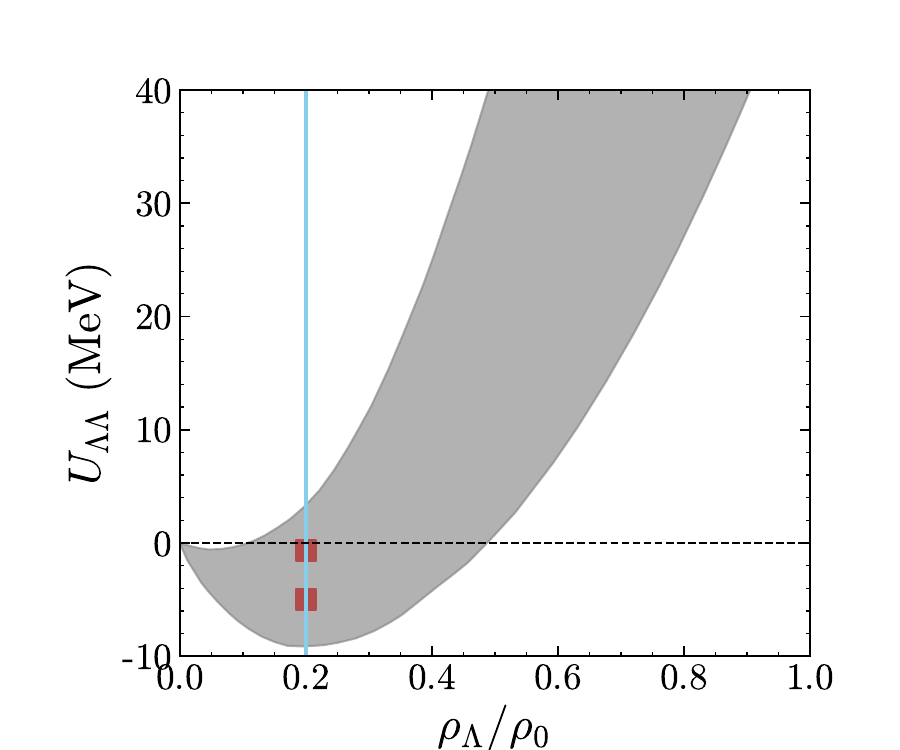}
\caption{Posterior $\Lambda\Lambda$ potential depth under +Astro+Nucl constraints, using SLy4+SLL4. The solid blue line represents 1/5 of saturation density of pure $\Lambda$ matter; the dashed line indicates $U_{\Lambda\Lambda} = 0$.
}
\label{fig:Upot}
\end{figure}

The analysis reveals that $\Lambda\Lambda$ parameters ($\lambda_0$--$\lambda_2$) are largely insensitive to variations in the underlying $NN$ and $\Lambda N$ interactions, since they are strongly constrained by the $\Lambda\Lambda$ potential depth $U_{\Lambda\Lambda}(\rho_0/5)$ (see Figure~\ref{fig:Post_Astro&Nucl}). In contrast, the $\Lambda\Lambda N$ parameters ($\lambda_3$, $\alpha$) exhibit strong sensitivity to $\Lambda N$ interactions. This arises because $\lambda_3$ governs the overall strength of the $\Lambda\Lambda N$ force, while $\alpha$ controls its density dependence. Consequently, both parameters directly influence the EOS stiffness and hence the maximum mass of hyperon stars. For example, the SLy4+YMR interaction, which generates the softest EOS, produces the largest posterior value of $\lambda_3$ and the smallest $\alpha$.

To compare parameter sets quantitatively, we evaluate the Bayes factors $\mathcal{B}$ (Table~\ref{tab:Post_NN&NY_combined_para}), defined as
\begin{equation}
\mathcal{B} = e^{\ln \mathcal{Z}_i - \ln \mathcal{Z}_j}
\end{equation}
where $\mathcal{Z}_i$ is the Bayesian evidence for model $i$, obtained by marginalizing the likelihood over the prior parameter space. Our analysis identifies SLy4+SLL4 as the most favored interaction, yielding the largest Bayes factor and providing the best simultaneous description of astrophysical and nuclear constraints.

Finally, Figure~\ref{fig:Upot} shows the posterior distribution of the $\Lambda$ potential depth $U_{\Lambda}^{\Lambda}$ in pure $\Lambda$ matter under +Astro+Nucl constraints. Using SLy4+SLL4, the results fully encompass the experimental values extracted from double-$\Lambda$ hypernuclei. A key feature is the sign change in $U_{\Lambda\Lambda}$ at densities $\sim 0.27^{+0.13}_{-0.14}\,\rho_{\Lambda}/\rho_0$, signaling a transition from attractive to repulsive $\Lambda\Lambda$ interactions. This density-dependent crossover strongly impacts both the EOS and particle fractions, as discussed below. 

\begin{table*}
	\centering
	\caption{
    Posterior hyperon star properties with or without $\Lambda\Lambda$N three-body forces (w.o.: without), based on $NN$+$\Lambda N$ and $NN$+$\Lambda N$+$\Lambda\Lambda$+$\Lambda\Lambda$N interactions at 68.3\% confidence interval, corresponding to Figure~\ref{fig:MR&MRrho&Density}.
    }	\label{tab:Post_YY&NYY_combined_MR}
\renewcommand\arraystretch{1.6}
	\setlength{\tabcolsep}{0.38cm}	
    \begin{tabular}{ccccccccccccc}
             \hline
            \hline
            $NN$ & $\Lambda \rm N$    &$\Lambda\Lambda$&$M_{\rm max}/M_{\odot}$    &$R_{\rm 2.0}\;\rm (km)$   &$\rho_{\rm core}\;\rm(fm^{-3})$ &$\rho_{\rm crit}\;\rm(fm^{-3})$          &$R_{1.4}\;\rm (km)$&$\Lambda_{1.4}$          \\
                        \hline
            \multirow{2}{*}{SLy4} & \multirow{2}{*}{SLL4}         & w.o. $\Lambda\Lambda$N & $1.98^{+0.24}_{-0.04}$  & $-$                   & $1.31^{+0.05}_{-0.05}$  & $0.52^{+0.00}_{-0.00}$  & $11.69^{+0.00}_{-0.00}$  & $299.41^{+0.04}_{-0.01}$ \\
            \multicolumn{2}{c}{}                                  & with $\Lambda\Lambda$N & $2.03^{+0.01}_{-0.03}$  & $9.82^{+0.09}_{-0.10}$  & $1.26^{+0.04}_{-0.03}$  & $0.52^{+0.00}_{-0.00}$  & $11.69^{+0.00}_{-0.00}$  & $299.41^{+0.04}_{-0.01}$  \\
            \cline{1-9}
            
            \multirow{2}{*}{SGI}  & \multirow{2}{*}{SLL4}         & w.o. $\Lambda\Lambda$N & $2.03^{+0.05}_{-0.07}$  & $10.15^{+0.22}_{-0.20}$ & $1.22^{+0.35}_{-0.21}$  & $0.32^{+0.00}_{-0.00}$  & $12.92^{+0.00}_{-0.00}$  & $577.09^{+24.63}_{-3.07}$ \\
            \multicolumn{2}{c}{}                                  & with $\Lambda\Lambda$N & $2.12^{+0.04}_{-0.06}$  & $10.45^{+0.29}_{-0.23}$ & $1.13^{+0.07}_{-0.06}$  & $0.32^{+0.00}_{-0.00}$  & $12.94^{+0.01}_{-0.01}$  & $577.09^{+24.63}_{-3.07}$ \\
            \cline{1-9}
            
            \multirow{2}{*}{SLy4} & \multirow{2}{*}{HP$\Lambda$2} & w.o. $\Lambda\Lambda$N & $1.89^{+0.02}_{-0.02}$  & $-$                   & $1.30^{+0.03}_{-0.05}$  & $0.52^{+0.00}_{-0.00}$  & $11.69^{+0.00}_{-0.00}$  & $295.44^{+0.00}_{-0.00}$\\
            \multicolumn{2}{c}{}                                  & with $\Lambda\Lambda$N & $2.00^{+0.01}_{-0.03}$  & $9.73^{+0.14}_{-0.13}$  & $1.23^{+0.04}_{-0.04}$  & $0.52^{+0.05}_{-0.05}$  & $11.69^{+0.02}_{-0.01}$  & $295.44^{+4.20}_{-1.69}$ \\
            \cline{1-9}
            
            \multirow{2}{*}{SKI3} & \multirow{2}{*}{SLL4}         & w.o. $\Lambda\Lambda$N & $2.01^{+0.05}_{-0.08}$  & $10.31^{+0.24}_{-0.27}$ & $1.20^{+0.08}_{-0.07}$  & $0.52^{+0.00}_{-0.00}$  & $13.65^{+0.02}_{-0.05}$  & $771.80^{+71.65}_{-8.70}$\\
            \multicolumn{2}{c}{}                                  & with $\Lambda\Lambda$N & $2.10^{+0.05}_{-0.7}$   & $10.62^{+0.36}_{-0.26}$ & $1.11^{+0.07}_{-0.08}$  & $0.52^{+0.00}_{-0.00}$  & $13.67^{+0.19}_{-0.01}$  & $771.80^{+71.65}_{-8.70}$\\
            \cline{1-9}
            
            \multirow{2}{*}{SLy4} & \multirow{2}{*}{YMR}          & w.o. $\Lambda\Lambda$N & $1.85^{+0.27}_{-0.05}$  & $-$                   & $1.42^{+0.06}_{-0.05}$  & $0.52^{+0.00}_{-0.00}$  & $11.56^{+0.07}_{-0.01}$  & $289.50^{+8.10}_{-3.07}$\\
            \multicolumn{2}{c}{}                                  & with $\Lambda\Lambda$N & $1.93^{+0.01}_{-0.01}$  & $-$                   & $1.22^{+0.02}_{-0.03}$  & $0.52^{+0.00}_{-0.00}$  & $11.56^{+0.07}_{-0.01}$  & $289.50^{+8.10}_{-3.07}$\\
            \hline
            \hline
	\end{tabular}
\end{table*}

\subsection{Evolution of baryon effective masses with density}

Beyond saturation properties, the density evolution of baryon effective masses plays a central role in determining the microscopic composition and macroscopic behavior of dense matter. In the SHF framework, effective masses encode the momentum dependence of the interaction and directly affect kinetic contributions to the pressure, chemical potentials, and hyperon population. In this section, we use the Bayesian-constrained interaction sets to elucidate how different hyperonic interactions modify effective masses and, in turn, the EOS and neutron star properties.

As shown in Fig.~\ref{fig: mass of baryon}, the effective masses of neutrons and protons decrease monotonically with increasing density and exhibit a clear isospin splitting in neutron-rich matter.
After the onset of $\Lambda$ hyperons, the inclusion of the $\Lambda N$ interaction significantly enhances the density dependence of nucleon effective masses. This modification lowers the neutron and proton effective masses more rapidly, thereby increasing their chemical potentials and favoring the conversion of nucleons into $\Lambda$ hyperons under chemical equilibrium. As a result, the nucleonic fractions are suppressed while the $\Lambda$ population grows rapidly with density, as shown below in the right panel of Fig.~\ref{fig:MR&MRrho&Density}. The enhanced hyperon fraction reduces the kinetic pressure contribution at high densities, leading to a pronounced softening of the EOS (Fig.~\ref{fig:EOS of YY+YYN}) and a substantial decrease in the maximum mass of hyperon-rich neutron stars (left panel of Fig.~\ref{fig:MR&MRrho&Density}).

When the $\Lambda\Lambda$ interaction is included, the effective masses of neutrons and protons are further reduced, which lowers their chemical potentials and counteracts the rapid buildup of the $\Lambda$ population at high densities. Consequently, a larger nucleonic component is retained in dense matter, suppressing hyperon dominance and increasing the pressure at a given energy density. This mechanism significantly stiffens the EOS compared to the $\Lambda N$–only case (Fig.~\ref{fig:EOS of YY+YYN}) and allows the maximum mass of hyperon-rich neutron stars to approach the observationally required value of $2\,M_{\odot}$ (left panel of Fig.~\ref{fig:MR&MRrho&Density}).

The inclusion of the $\Lambda\Lambda N$ three-body interaction introduces an additional and qualitatively important effect on the effective-mass evolution. While the nucleon effective masses are further suppressed, the effective mass of the $\Lambda$ hyperon increases with density. This interplay disfavors the early dominance of hyperons by increasing their kinetic contribution and delaying their rapid accumulation at high densities. The resulting enhancement of repulsion at supranuclear densities leads to a further stiffening of the EOS and stabilizes massive hyperon-rich neutron stars within the Bayesian-constrained parameter space, as seen in the left panel of Fig.~\ref{fig:MR&MRrho&Density}.

Overall, the density dependence of effective masses provides a clear microscopic interpretation of how different hyperonic interactions shape the EOS and neutron star observables. In particular, the repulsive effects associated with $\Lambda\Lambda$ and $\Lambda\Lambda N$ interactions manifest not only at the level of the energy density but also through their impact on effective masses, particle fractions, and chemical equilibrium. These trends are robust across the posterior distributions obtained in our Bayesian analysis and highlight the essential role of $YY$ and hyperonic three-body forces in resolving the hyperon puzzle within the SHF framework.

\subsection{Impact on the EOS and Neutron Star Observables}

Table~\ref{tab:Post_YY&NYY_combined_MR} summarizes the preferred hyperon star properties constrained by +Astro+Nucl. The maximum masses range from $1.93^{+0.04}_{-0.02}$ to $2.10^{+0.13}_{-0.07}\,M_{\odot}$, with corresponding radii of $9.82^{+0.17}_{-0.16}$ to $10.62^{+0.54}_{-1.11}$ km for $2\,M_{\odot}$ stars. The central densities span $1.13^{+0.13}_{-0.09}$ to $1.33^{+0.04}_{-0.04}\,\mathrm{fm}^{-3}$. 
Neither $\Lambda\Lambda$ nor $\Lambda\Lambda N$ interactions significantly alter the critical density $\rho_{\rm crit}$ for $\Lambda$ hyperon onset.
Note that, to clarify the role of $\Lambda\Lambda N$ forces, we perform two complementary Bayesian analyses: (i) a baseline model with only $\Lambda\Lambda$ interactions ($\lambda_0$-$\lambda_2$), and (ii) an extended model incorporating $\Lambda\Lambda N$ interactions ($\lambda_3$, $\alpha$). Figures~\ref{fig:EOS of YY+YYN} and~\ref{fig:MR&MRrho&Density} show the corresponding EOSs and hyperon star properties.

As seen in Figure~\ref{fig:EOS of YY+YYN}, $\Lambda\Lambda$ interactions stiffen the EOS at high densities ($\rho \gtrsim 0.60\,\rho_0$), while softening it in the intermediate range ($0.52$--$0.60\,\rho_0$). This behavior arises from the density-dependent sign change of $U_{\Lambda\Lambda}$: attractive at low $\Lambda$ densities but repulsive at high densities. The inclusion of $\Lambda\Lambda N$ forces further stiffens the EOS, though it never exceeds the stiffness of neutron star matter.

The impact on stellar properties is quantified in Table~\ref{tab:Post_YY&NYY_combined_MR} and Figure~\ref{fig:MR&MRrho&Density}. 
For the SLy4+SLL4 interaction, the appearance of $\Lambda$ hyperons reduces the maximum mass by $\sim25\%$, from $2.05\,M_{\odot}$ to $1.54\,M_{\odot}$, exemplifying the hyperon puzzle (see Table~\ref{tab:MR}). 
With $\Lambda\Lambda$ interactions included, the maximum mass increases to $1.98^{+0.24}_{-0.04}\,M_{\odot}$, representing a $\sim22\%$ enhancement, while the $\Lambda$ fraction is significantly reduced.
Incorporation of $\Lambda\Lambda N$ interactions yields a further increase to $2.03^{+0.01}_{-0.03}\,M_{\odot}$, with central density reduced from $1.31^{+0.05}_{-0.05}$ to $1.26^{+0.04}_{-0.03}\,\mathrm{fm}^{-3}$. 
Across all $NN$+$\Lambda N$ combinations, $\Lambda\Lambda N$ forces enhance the maximum mass by $2.5$--$5.8\%$, though they leave the radius and tidal deformability of $1.4\,M_{\odot}$ stars essentially unchanged.

\section{Summary}

In this work, we combine hypernuclear experimental data with the latest multi-messenger astronomical observations to perform a Bayesian analysis of the $\Lambda\Lambda$ and $\Lambda\Lambda N$ interaction parameters within the Skyrme-type effective interaction framework. 

Using Hartree-Fock calculations with Skyrme-type $\Lambda\Lambda$ and $\Lambda\Lambda N$ interactions, the five parameters $\lambda_0$, $\lambda_1$, $\lambda_2$, $\lambda_3$, and $\alpha$ capture distinct physical contributions. The interplay between the attractive local term ($\lambda_0$), the repulsive momentum-dependent terms ($\lambda_1$ and $\lambda_2$), and the repulsive density-dependent three-body terms ($\lambda_3$ and $\alpha$) shapes the effective hyperon interactions. These parameters are tightly constrained by the combined nuclear and astrophysical data, providing a comprehensive description of dense matter relevant for hyperon star modeling. 
Several interesting observations can be made:
\begin{itemize}
    \item \textbf{$\lambda_0$:} This parameter represents the \emph{local}, \emph{momentum-independent} component of the $\Lambda\Lambda$ interaction. It encodes the short-range attractive or repulsive effects independent of particle momenta. Our Bayesian analysis shows that $\lambda_0$ is tightly constrained and predominantly \emph{attractive}, as indicated by the negative peak in its posterior distribution. It sets the baseline strength of the two-body $\Lambda\Lambda$ force.

    \item \textbf{$\lambda_1$ and $\lambda_2$:} These parameters correspond to \emph{nonlocal}, \emph{momentum-dependent} contributions of the $\Lambda\Lambda$ interaction. Through gradient and kinetic energy terms, they introduce repulsion that increases with density and momentum. Posterior distributions favor positive values for $\lambda_1$ and $\lambda_2$, indicating that these terms counterbalance the attraction at high densities relevant for neutron star interiors.

    \item \textbf{$\lambda_3$:} This parameter controls the strength of the \emph{density-dependent} component of the three-body $\Lambda\Lambda N$ interaction. It contributes a repulsive term that grows with baryon density, stiffening the EOS and enabling support for neutron stars with $M>2\,M_{\odot}$. Our results prefer larger $\lambda_3$ values, highlighting the critical role of three-body repulsion in resolving the hyperon puzzle.

    \item \textbf{$\alpha$:} The parameter $\alpha$ modulates the \emph{power-law density dependence} of the three-body interaction. It determines how rapidly the repulsive effect grows with density. Astrophysical constraints tend to favor smaller $\alpha$, implying a moderate density dependence that remains consistent with neutron star radius and tidal deformability measurements.
\end{itemize}

In the current Bayesian analysis, we find that the $\Lambda\Lambda$ interaction provides a weakly attractive contribution at low $\Lambda$ densities, consistent with the negative potential depth extracted from hypernuclear experiments. This enhances the $\Lambda$ hyperon fraction and slightly softens the EOS. At higher $\Lambda$ densities, the potential depth becomes positive, suppressing the $\Lambda$ fraction, stiffening the EOS, and increasing the maximum mass of hyperon stars.
The $\Lambda\Lambda N$ interaction further affects the properties of massive hyperon stars. Across all five $NN$+$\Lambda N$ combinations studied, including the $\Lambda\Lambda N$ interaction consistently enhances the maximum mass by 2.5\%–5.8\%, while leaving the radius and tidal deformability of a $1.4\,M_{\odot}$ hyperon star essentially unchanged.

As the first Bayesian study of hyperon stars using SHF interactions constrained by both $\Lambda\Lambda$ hypernuclei experiments and multi-messenger astrophysical data, several caveats remain. 
First, the $\Lambda\Lambda$ potential depth at one-fifth of nuclear saturation density is still subject to significant uncertainty. Future large-scale hypernuclear experiments are expected to reduce this uncertainty, thereby enabling more precise predictions for the properties of hyperon-rich neutron stars.
Moreover, our analysis is restricted to $\Lambda$ hyperons, whereas neutron star matter at high densities may also contain $\Sigma$ and $\Xi$ hyperons. The inclusion of these additional strangeness carriers is expected to modify the EOS in two main ways. First, the onset of hyperons would occur at lower densities, leading to an earlier softening of the EOS. Second, the additional hyperon species would introduce new interaction channels (e.g., $\Xi N$, $\Xi\Xi$, etc.) that could either soften or stiffen the EOS depending on their underlying strengths and signs. Currently, these interactions are poorly constrained by experiments, and their Skyrme-type parametrizations are not well established.
Nevertheless, the repulsive $\Lambda\Lambda$ and $\Lambda\Lambda N$ interactions constrained in this work would play a similar role in counterbalancing the overall softening induced by hyperons, by providing repulsion at high densities. Future work extending the Skyrme functional to include $\Sigma$ and $\Xi$ hyperons, as well as their two- and three-body interactions, will be essential for a complete description of strangeness-rich matter in neutron stars.

\section*{Acknowledgments}
The work is supported by the National Natural Science Foundation of China (grant Nos. 12273028, 12494572, 12475149).

\section*{Data Availability}
The data that support the findings of this article are openly available [1,2,3]. 



\bibliographystyle{mnras}
\bibliography{yy} 


\onecolumn

\appendix

\section{Nucleon--Nucleon Hamiltonian from Detailed Interactions}
\label{Appendix:A2}

\subsection{Kinetic Term}
Assuming nucleons move independently in a Skyrme-type potential, the expectation value of the total energy reads
\begin{eqnarray}
E &=& \int H(\bm{r})\,d\bm{r} \nonumber\\
  &=& \langle \phi | T + V | \phi \rangle \nonumber\\
  &=& \sum_{i}\left\langle i \left| \frac{\hat{\bm{p}}^2}{2m} \right| i \right\rangle
  + \frac{1}{2}\langle ij|\tilde{v}_{12}|ij\rangle
  + \frac{1}{6}\langle ijk|\tilde{v}_{123}|ijk\rangle .
\end{eqnarray}

For the kinetic contribution one obtains
\begin{eqnarray}
E_T &=& \sum_i \left\langle i \left| \frac{\hat{\bm{p}}^2}{2m} \right| i \right\rangle \nonumber\\
    &=& \int d\bm{r}\,\sum_q \frac{\hbar^2}{2m_q}\tau_q ,
\end{eqnarray}
so that the Hamiltonian density of the kinetic term is
\begin{eqnarray}
\mathcal{H}^{\text{kin}}_{NN} = \sum_q \frac{\hbar^2}{2m_q}\tau_q .
\end{eqnarray}

\subsection{S-Wave Central Interaction: \texorpdfstring{$v_0(NN)$}{v0(NN)}}
The zero-range central interaction is
\begin{eqnarray}
v_0(NN) = t_0(1+x_0 P^{\sigma}) \delta(\bm{r}) .
\end{eqnarray}
The corresponding contribution to the energy reads
\begin{eqnarray}\label{eq:E_v0_NN}
E^{v_0}_{NN}
&=& \frac{1}{2}\sum_{ij}\langle ij|v_0(1-P^M P^{\sigma} P^{\tau})|ij\rangle \nonumber\\
&=& \int d\bm{r}\,\frac{t_0}{2}\Big[(1+\tfrac{1}{2}x_0)\rho^2 - (\tfrac{1}{2}+x_0)\sum_q \rho_q^2 \Big] .
\end{eqnarray}
Thus, the Hamiltonian density is
\begin{eqnarray}\label{eq:H_v0_NN}
\mathcal{H}^{v_0}_{NN} = \frac{t_0}{2}\Big[(1+\tfrac{1}{2}x_0)\rho^2 - (\tfrac{1}{2}+x_0)\sum_q \rho_q^2 \Big] .
\end{eqnarray}

\subsection{S-Wave Momentum-Dependent Interaction: \texorpdfstring{$v_1(NN)$}{v1(NN)}}
The momentum-dependent S-wave interaction is
\begin{eqnarray}
v_1(NN) = \tfrac{1}{2}t_1(1+x_1P^{\sigma})[\bm{k}^{\prime 2}\delta(\bm{r}) + \delta(\bm{r})\bm{k}^2] .
\end{eqnarray}
After decomposing into Hartree and Fock terms and performing the standard derivations, one obtains the Hamiltonian density
\begin{eqnarray}\label{eq:H_v1_NN}
\mathcal{H}^{v_1}_{NN} &=& 
\frac{1}{16}t_1(1+\tfrac{1}{2}x_1)\Big[ 4\rho\tau + 3(\nabla\rho)^2 \Big] \nonumber\\
&\quad&- \frac{1}{16}t_1(1+\tfrac{1}{2}x_1)\sum_q \Big[ 4\rho_q\tau_q + 3(\nabla\rho_q)^2 \Big] \nonumber\\
&\quad&- \frac{1}{16}t_1x_1\bm{J}^2 + \frac{1}{16}t_1x_1\sum_q \bm{J}_q^2 .
\end{eqnarray}

\subsection{P-Wave Central Interaction: \texorpdfstring{$v_2(NN)$}{v2(NN)}}
The P-wave interaction takes the form
\begin{eqnarray}
v_2(NN) = t_2(1+x_2P^{\sigma}) \bm{k}'\cdot\delta(\bm{r})\,\bm{k} .
\end{eqnarray}
Following similar steps as above, the Hamiltonian density becomes
\begin{eqnarray}\label{eq:H_v2_NN}
\mathcal{H}^{v_2}_{NN} &=&
\frac{1}{16}t_2(1+\tfrac{1}{2}x_2)\Big[ 4\rho\tau - (\nabla\rho)^2 \Big] \nonumber\\
&\quad&+ \frac{1}{16}t_2(\tfrac{1}{2}+x_2)\sum_q \Big[ 4\rho_q\tau_q - (\nabla\rho_q)^2 \Big] \nonumber\\
&\quad&- \frac{1}{16}t_2x_2\bm{J}^2 - \frac{1}{16}t_2\sum_q \bm{J}_q^2 .
\end{eqnarray}

\subsection{Density-Dependent Two-Body Force}
For the density-dependent term in the S-wave channel, one has
\begin{eqnarray}
v_{den3}(NN) = \tfrac{1}{6}t_{3i}(1+x_{3i}P^{\sigma})\rho^{\alpha_i}(\bm{R})\delta(\bm{r}) .
\end{eqnarray}
The Hamiltonian density contribution reads
\begin{eqnarray}\label{eq:H_den3_NN}
\mathcal{H}^{den3}_{NN} =
\frac{1}{12}t_{3i}\rho^{\alpha_i}\Big[(1+\tfrac{1}{2}x_{3i})\rho^2 - (\tfrac{1}{2}+x_{3i})\sum_q \rho_q^2 \Big] .
\end{eqnarray}

\subsection{Summary of the $NN$ Hamiltonian}
Collecting all contributions, the total Hamiltonian density of nucleon--nucleon interactions in uniform infinite nuclear matter is
\begin{eqnarray}
\mathcal{H}_{NN} &=& \mathcal{H}^{kin}_{NN} + \mathcal{H}^{v_0}_{NN} + \mathcal{H}^{v_1}_{NN} + \mathcal{H}^{v_2}_{NN} + \mathcal{H}^{den3}_{NN} \nonumber\\
&=& \sum_{q=n,p}\frac{\hbar^2}{2m_q}\tau_q \nonumber\\
&\quad&+ \rho(\tau_n+\tau_p)\Big[ \tfrac{t_1}{4}(1+\tfrac{x_1}{2}) + \tfrac{t_2}{4}(1+\tfrac{x_2}{2}) \Big] \nonumber\\
&\quad&+ \sum_{q=n,p}\tau_q\rho_q \Big[ -\tfrac{t_1}{4}(\tfrac{1}{2}+x_1) + \tfrac{t_2}{4}(\tfrac{1}{2}+x_2) \Big] \nonumber\\
&\quad&+ \tfrac{t_0}{2}\Big[(1+\tfrac{x_0}{2})\rho^2 - (\tfrac{1}{2}+x_0)(\rho_n^2+\rho_p^2)\Big] \nonumber\\
&\quad&+ \tfrac{t_3}{12}\Big[(1+\tfrac{x_3}{2})\rho^2 - (\tfrac{1}{2}+x_3)(\rho_n^2+\rho_p^2)\Big]\rho^{\epsilon}.
\end{eqnarray}

\section{$\Lambda N$ Hamiltonian from Detailed Interactions}
\label{Appendix:A3}

In this section we derive the Hamiltonian density for $\Lambda$--nucleon ($\Lambda N$) interactions within the Skyrme framework, following the same procedure as in Appendix~\ref{Appendix:A2}. Since $\Lambda$ hyperons are distinguishable from nucleons, exchange terms differ from the $NN$ case.

\subsection{Kinetic Term of the $\Lambda$}
The kinetic energy density of the $\Lambda$ hyperon is
\begin{eqnarray}
\mathcal{H}^{\text{kin}}_{\Lambda} = \frac{\hbar^2}{2m_{\Lambda}}\tau_{\Lambda} ,
\end{eqnarray}
where $m_{\Lambda}$ is the $\Lambda$ mass, and $\tau_{\Lambda}$ the kinetic energy density.

\subsection{S-Wave Central Interaction: \texorpdfstring{$u_0(\Lambda N)$}{u0(LN)}}
The zero-range central part of the $\Lambda N$ force is written as
\begin{eqnarray}
v_0(\Lambda N) = u_0(1+y_0 P^{\sigma})\delta(\bm{r}) .
\end{eqnarray}
The contribution to the Hamiltonian density is
\begin{eqnarray}\label{eq:H_v0_LN}
\mathcal{H}^{v_0}_{\Lambda N} = \frac{u_0}{2}\Big[(1+\tfrac{1}{2}y_0)\rho_{\Lambda}\rho_N 
- (\tfrac{1}{2}+y_0)\sum_q \rho_{\Lambda}\rho_q \Big] ,
\end{eqnarray}
where $\rho_N = \rho_n + \rho_p$ denotes the total nucleon density.

\subsection{Momentum-Dependent Terms: \texorpdfstring{$u_1(\Lambda N)$}{u1(LN)} and \texorpdfstring{$u_2(\Lambda N)$}{u2(LN)}}
The momentum-dependent interactions are
\begin{eqnarray}
v_1(\Lambda N) &=& \tfrac{1}{2}u_1(1+y_1 P^{\sigma})[\bm{k}'^2\delta(\bm{r}) + \delta(\bm{r})\bm{k}^2] , \\
v_2(\Lambda N) &=& u_2(1+y_2 P^{\sigma})\bm{k}'\cdot\delta(\bm{r})\,\bm{k} .
\end{eqnarray}

The resulting Hamiltonian density is
\begin{eqnarray}\label{eq:H_v12_LN}
\mathcal{H}^{v_1+v_2}_{\Lambda N} &=&
\frac{1}{8}(u_1+u_2)(\rho_N\tau_{\Lambda} + \rho_{\Lambda}\tau_N) \nonumber\\
&\quad&+ \frac{1}{8}(u_1- u_2)\sum_q \rho_q\tau_{\Lambda} 
+ \frac{1}{8}(u_1- u_2)\rho_{\Lambda}\sum_q \tau_q \nonumber\\
&\quad&- \frac{1}{8}(3u_1 - u_2)(\bm{j}_N\cdot\bm{j}_{\Lambda}) ,
\end{eqnarray}
where $\bm{j}_N$ and $\bm{j}_{\Lambda}$ are the current densities of nucleons and $\Lambda$, respectively.

\subsection{Density-Dependent Interaction: \texorpdfstring{$u_3'(\Lambda N)$}{u3'(LN)}}
The $\Lambda N$ density-dependent interaction takes the form
\begin{eqnarray}
v_{den3}(\Lambda N) = \tfrac{3}{8}u_3'(1+y_3P^{\sigma})\rho_N^{\gamma}\,\delta(\bm{r}) .
\end{eqnarray}
This leads to the Hamiltonian density
\begin{eqnarray}\label{eq:H_den3_LN}
\mathcal{H}^{den3}_{\Lambda N} = \frac{3}{8}u_3'\rho_N^{\gamma}\Big[(1+\tfrac{1}{2}y_3)\rho_{\Lambda}\rho_N
- (\tfrac{1}{2}+y_3)\sum_q \rho_{\Lambda}\rho_q \Big] .
\end{eqnarray}

\subsection{Summary of the $\Lambda N$ Hamiltonian}
Summing all contributions, the total $\Lambda N$ Hamiltonian density reads
\begin{eqnarray}
\mathcal{H}_{\Lambda N} &=& \mathcal{H}^{\text{kin}}_{\Lambda} + \mathcal{H}^{v_0}_{\Lambda N} + \mathcal{H}^{v_1+v_2}_{\Lambda N} + \mathcal{H}^{den3}_{\Lambda N} \nonumber\\
&=& \frac{\hbar^2}{2m_{\Lambda}}\tau_{\Lambda} \nonumber\\
&\quad&+ \frac{u_0}{2}\Big[(1+\tfrac{1}{2}y_0)\rho_{\Lambda}\rho_N 
- (\tfrac{1}{2}+y_0)\sum_q \rho_{\Lambda}\rho_q \Big] \nonumber\\
&\quad&+ \frac{1}{8}(u_1+u_2)(\rho_N\tau_{\Lambda} + \rho_{\Lambda}\tau_N) \nonumber\\
&\quad&+ \frac{1}{8}(u_1- u_2)\sum_q \rho_q\tau_{\Lambda} 
+ \frac{1}{8}(u_1- u_2)\rho_{\Lambda}\sum_q \tau_q \nonumber\\
&\quad&- \frac{1}{8}(3u_1 - u_2)(\bm{j}_N\cdot\bm{j}_{\Lambda}) \nonumber\\
&\quad&+ \frac{3}{8}u_3'\rho_N^{\gamma}\Big[(1+\tfrac{1}{2}y_3)\rho_{\Lambda}\rho_N
- (\tfrac{1}{2}+y_3)\sum_q \rho_{\Lambda}\rho_q \Big] .
\end{eqnarray}

\noindent
Here $u_0$, $u_1$, $u_2$, and $u_3'$ denote the central, momentum-dependent, and density-dependent $\Lambda N$ Skyrme parameters, while $y_i$ are their spin-exchange factors. These terms respectively capture local attraction, repulsive momentum dependence, and medium modifications essential for describing hypernuclei and hyperonic matter in neutron stars.

\section{$\Lambda\Lambda$ Hamiltonian from Detailed Interactions}
\label{Appendix:A4}

We now construct the Hamiltonian density for the $\Lambda\Lambda$ interaction in the Skyrme framework.  
Unlike the $\Lambda N$ case, both particles are identical fermions, and full antisymmetrization must be applied.

\subsection{Kinetic Term of the $\Lambda$}
The kinetic energy density is already given in Appendix~\ref{Appendix:A3} as
\begin{eqnarray}
\mathcal{H}^{\text{kin}}_{\Lambda} = \frac{\hbar^2}{2m_{\Lambda}}\tau_{\Lambda} .
\end{eqnarray}
This term is included here for completeness.

\subsection{S-Wave Central Interaction: \texorpdfstring{$\lambda_0(\Lambda\Lambda)$}{lambda0(LL)}}
The zero-range central force is
\begin{eqnarray}
v_0(\Lambda\Lambda) = \lambda_0(1 - P^{\sigma})\delta(\bm{r}) ,
\end{eqnarray}
where the antisymmetry under particle exchange enforces the $1-P^\sigma$ structure.  

The Hamiltonian density becomes
\begin{eqnarray}
\mathcal{H}^{v_0}_{\Lambda\Lambda} = \frac{\lambda_0}{4}\rho_{\Lambda}^2 .
\end{eqnarray}

\subsection{Momentum-Dependent Terms: \texorpdfstring{$\lambda_1$}{lambda1} and \texorpdfstring{$\lambda_2$}{lambda2}}
The momentum-dependent forces are
\begin{eqnarray}
v_1(\Lambda\Lambda) &=& \tfrac{1}{2}\lambda_1\left[\bm{k}'^2\delta(\bm{r}) + \delta(\bm{r})\bm{k}^2\right] , \\
v_2(\Lambda\Lambda) &=& \lambda_2 \bm{k}'\cdot\delta(\bm{r})\,\bm{k} .
\end{eqnarray}

The contribution to the Hamiltonian density is
\begin{eqnarray}
\mathcal{H}^{v_1+v_2}_{\Lambda\Lambda} &=&
\frac{1}{8}(\lambda_1+3\lambda_2)\rho_{\Lambda}\tau_{\Lambda}
- \frac{1}{8}(\lambda_1- \lambda_2)\bm{j}_{\Lambda}^2 .
\end{eqnarray}

\subsection{Density-Dependent Interaction: \texorpdfstring{$\lambda_3$}{lambda3}}
The $\Lambda\Lambda$ density-dependent term is modeled as
\begin{eqnarray}
v_{den3}(\Lambda\Lambda) = \tfrac{1}{6}\lambda_3 \rho_{\Lambda}^{\alpha} \delta(\bm{r}) ,
\end{eqnarray}
leading to the Hamiltonian density
\begin{eqnarray}
\mathcal{H}^{den3}_{\Lambda\Lambda} = \frac{1}{24}\lambda_3 \rho_{\Lambda}^{\alpha+2} .
\end{eqnarray}
Here $\alpha$ is an empirical exponent controlling the medium dependence of the interaction.

\subsection{Summary of the $\Lambda\Lambda$ Hamiltonian}
Collecting all terms, the total Hamiltonian density for the $\Lambda\Lambda$ interaction reads
\begin{eqnarray}
\mathcal{H}_{\Lambda\Lambda} &=& \frac{\hbar^2}{2m_{\Lambda}}\tau_{\Lambda} 
+ \frac{\lambda_0}{4}\rho_{\Lambda}^2 \nonumber\\
&\quad&+ \frac{1}{8}(\lambda_1+3\lambda_2)\rho_{\Lambda}\tau_{\Lambda}
- \frac{1}{8}(\lambda_1- \lambda_2)\bm{j}_{\Lambda}^2 \nonumber\\
&\quad&+ \frac{1}{24}\lambda_3 \rho_{\Lambda}^{\alpha+2} .
\end{eqnarray}

\noindent
Here $\lambda_0$ represents the local, momentum-independent central interaction; $\lambda_1$ and $\lambda_2$ parameterize the momentum-dependent components; $\lambda_3$ and $\alpha$ govern the density-dependent repulsion.  
Together, these parameters control the balance between attraction at low densities (crucial for hypernuclear binding) and repulsion at high densities (necessary for supporting $2M_\odot$ neutron stars).

\section{Posterior Distributions with Different $NN$ and \texorpdfstring{$\Lambda$}{Λ}N Combinations}\label{Appendix:A1}

We present the posterior distributions of the $\Lambda\Lambda$ interaction parameters ($\lambda_0$, $\lambda_1$, $\lambda_2$) and the $\Lambda\Lambda N$ three-body parameters ($\lambda_3$, $\alpha$) at the 68.3\% confidence level. The results corresponding to the SLy4+SLL4 interaction set are summarized in Table~\ref{tab:Post_NN&NY_combined_para}.

Table \ref{tab:Post_NN&NY_combined_MR} presents the posterior properties of hyperon stars obtained from our Bayesian analysis under combined astrophysical and nuclear constraints (+Astro+Nucl). This comprehensive table summarizes key observables—including maximum mass (\(M_{\text{max}}\)), radius at \(2\,M_{\odot}\) (\(R_{2.0}\)), central density (\(\rho_{\text{core}}\)), hyperon onset density (\(\rho_{\text{crit}}\)), and properties of \(1.4\,M_{\odot}\) stars—for all five \(NN\)+\(\Lambda N\) interaction combinations studied. These results complement the posterior parameter distributions shown in Figure \ref{fig:PDF of 1D YY+YYN} and provide the complete set of constrained hyperon star properties referenced throughout Section \ref{sec:result}.

\begin{figure*}
\centering
\includegraphics[width=0.43\textwidth]{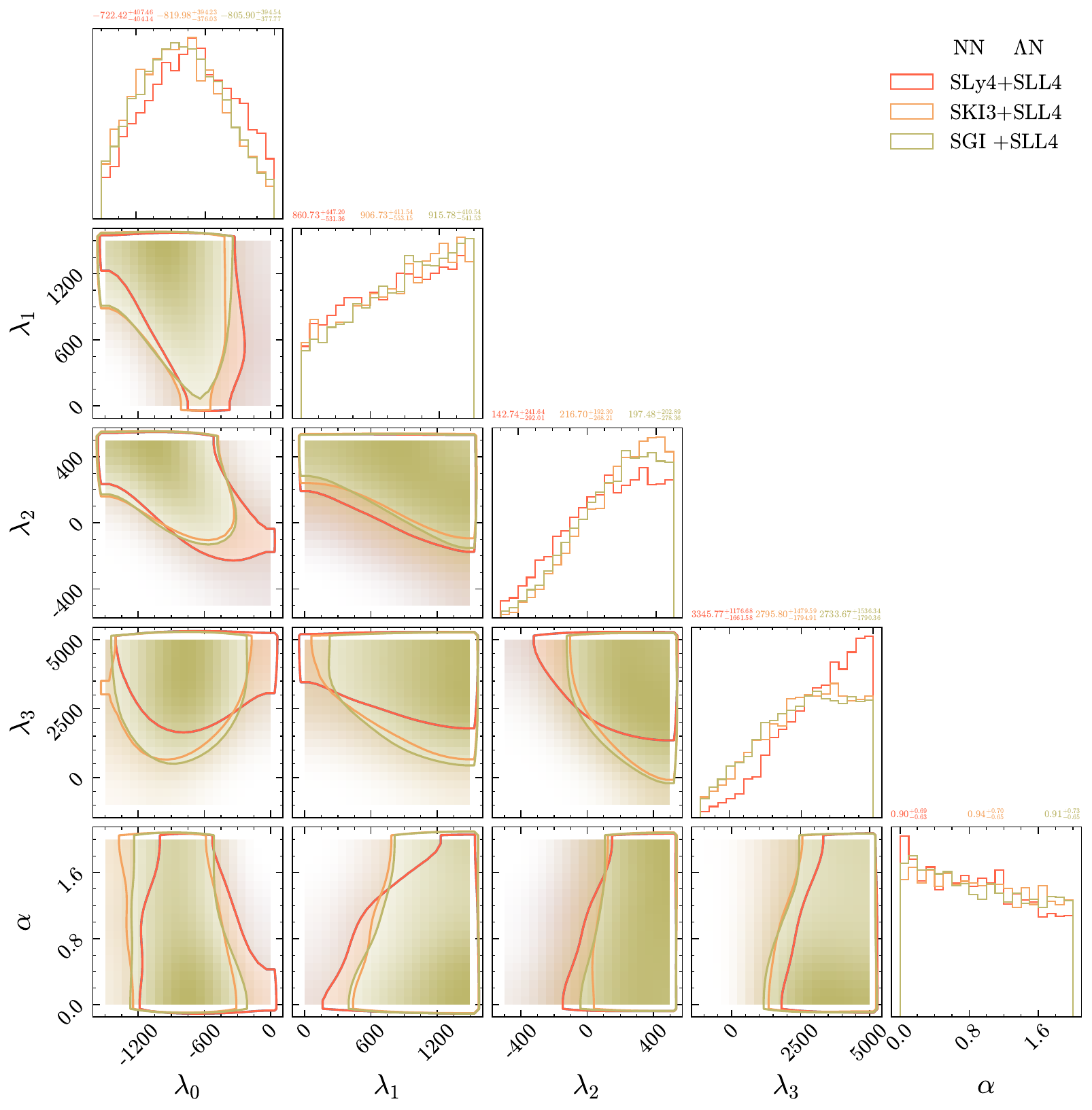}
\includegraphics[width=0.43\textwidth]{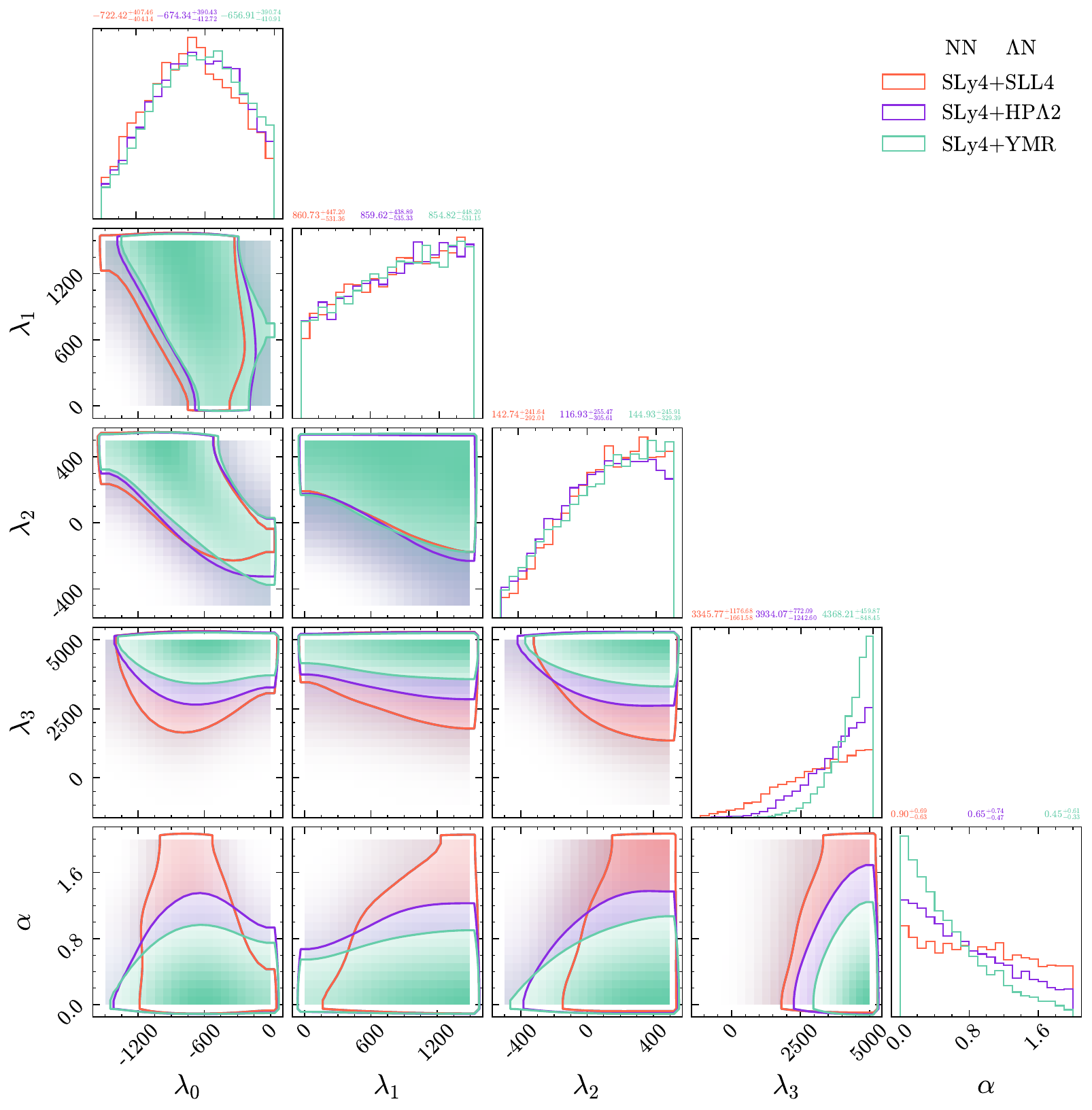}
\caption{Posterior distributions of $\lambda_0$, $\lambda_1$, $\lambda_2$, $\lambda_3$, and $\alpha$ for different $NN$ interactions under combined +Astro+Nucl constraints. The left panel corresponds to variations in $NN$ forces, while the right panel shows the effects of different $\Lambda N$ interactions.}
\label{fig:NN_Total}
\end{figure*}

\begin{table*}
	\centering
	\caption{Posterior properties of hyperon stars with different $NN$+$\Lambda N$ interactions (SLy4+SLL4, SGI+SLL4, etc.), constrained by +Astro+Nucl. $M_{\rm max}$: maximum mass; $R_{2.0}$: radius at 2 $M_\odot$; $\rho_{\rm core}$: maximum central density; $\rho_{\rm crit}$: $\Lambda$ appearance threshold; $R_{1.4}$ and $\Lambda_{1.4}$: radius and tidal deformability of 1.4 $M_\odot$ stars.
    }	\label{tab:Post_NN&NY_combined_MR}
\renewcommand\arraystretch{1.6}
	\setlength{\tabcolsep}{0.5cm}	\begin{tabular}{cccccccccccc} 
                 \hline
            \hline
            $NN$ & $\Lambda \rm N$ & $M_{\rm max}/M_{\odot}$      &$R_{2.0}\;\rm (km)$      &$\rho_{\rm core}\;\rm(fm^{-3})$ &$\rho_{\rm crit}\;\rm(fm^{-3})$ &$R_{1.4}\;\rm (km)$ &$\Lambda_{1.4}$\\
            \hline
		  SLy4 & SLL4          & $2.03^{+0.09}_{-0.01}$   &$9.88^{+0.22}_{-0.63}$   &$1.26^{+0.05}_{-0.03}$  &$0.52^{+0.00}_{-0.00}$ &$11.70^{+0.00}_{-0.00}$ &$299.41^{+0.04}_{-0.01}$    \\
            SGI & SLL4           & $2.10^{+0.13}_{-0.07}$   &$10.47^{+0.48}_{-1.21}$  &$1.15^{+0.13}_{-0.07}$  &$0.32^{+0.00}_{-0.00}$ &$12.98^{+0.04}_{-0.01}$ &$577.09^{+24.63}_{-3.07}$     \\
            SLy4 & HP$\Lambda$2  & $2.00^{+0.06}_{-0.01}$   &$9.82^{+0.17}_{-0.16}$   &$1.25^{+0.14}_{-0.03}$  &$0.52^{+0.00}_{-0.00}$ &$11.69^{+0.02}_{-0.01}$ &$295.44^{+4.20}_{-1.69}$  \\
		  SKI3 & SLL4          & $2.08^{+0.13}_{-0.6}$    &$10.62^{+0.54}_{-1.11}$  &$1.13^{+0.13}_{-0.09}$  &$0.52^{+0.00}_{-0.00}$ &$13.69^{+0.14}_{-0.02}$ &$771.80^{+71.65}_{-8.70}$   \\
            SLy4 & YMR           & $1.93^{+0.04}_{-0.02}$   &$-$                    &$1.33^{+0.04}_{-0.04}$  &$0.52^{+0.00}_{-0.00}$ &$11.66^{+0.04}_{-0.01}$ &$289.50^{+8.10}_{-3.07}$  \\
                         \hline
            \hline
	\end{tabular}
\end{table*}

\bsp	
\label{lastpage}
\end{document}